\documentclass{aa}

\usepackage{graphicx, comment}
\usepackage{txfonts}
\usepackage[colorlinks,urlcolor=blue,citecolor=blue,linkcolor=blue]{hyperref}

\bibpunct{(}{)}{;}{a}{}{,}

\usepackage{xcolor}
\usepackage{xspace}

\def\est#1{\textcolor{gray}{#1}\xspace}
\newcommand{\frbpoppy}{\texttt{frbpoppy}\xspace}
\newcommand{\FWHM}{\textsc{FWHM}\xspace}
\newcommand{\SEFD}{\textsc{SEFD}\xspace}
\def\pop#1{\emph{#1}\xspace}
\def\survey#1{`#1'\xspace}
\hypersetup{
    colorlinks,
    linkcolor={red!60!black},
    citecolor={blue!60!black},
    urlcolor={blue!80!black}
}

\usepackage[frozencache]{minted}

\usepackage[normalem]{ulem}
\makeatletter
\renewcommand*\aa@pageof{, page \thepage{} of \pageref*{LastPage}}
\makeatother

\begin{document}

\title{Synthesizing the intrinsic FRB population using  \frbpoppy}

\author{D.W. Gardenier
 \inst{1, 2}\fnmsep\thanks{\email{gardenier@astron.nl}}
 \and
 J. van Leeuwen\inst{1, 2}
 \and
 L. Connor\inst{2}
 \and
 E. Petroff\inst{2}
}

\institute{ASTRON - the Netherlands Institute for Radio Astronomy, Oude Hoogeveensedijk 4, 7991 PD, Dwingeloo, The Netherlands
 \and
 Anton Pannekoek Institute for Astronomy, University of Amsterdam, Science Park 904, 1098 XH Amsterdam, The Netherlands
}

\date{Received TODO; accepted TODO}

\abstract
{
Fast Radio Bursts (FRBs) are radio transients of an unknown origin.
Naturally, we are curious as to their nature.
Enough FRBs have been detected for a statistical approach to parts of this challenge to be feasible.
To understand the crucial link between detected FRBs and the underlying FRB source classes
we perform FRB population synthesis, to determine how the underlying population behaves.
The Python package we developed for this synthesis, \frbpoppy, is open source and freely available.

}
{Our goal is to determine the current best fit FRB population model.
Our secondary aim is to provide an easy-to-use tool for simulating and understanding  FRB detections.
  It can compare  surveys, or inform us of the intrinsic FRB population.
}
{\frbpoppy simulates intrinsic FRB populations and the surveys that find them, to produce virtual observed populations. These resulting populations can then be compared with real data, allowing constrains to be placed on underlying physics and selection effects. }
{We are  able to replicate real Parkes and ASKAP FRB surveys, in terms of both detection rates and distributions
  observed. We also show the effect of beam patterns on the observed dispersion measure (DM) distributions.
  We compare four types of source  models. The "Complex" model, featuring a range of luminosities, pulse widths and
  spectral indices, reproduces current detections  best.}
{Using \frbpoppy, an open-source FRB population synthesis package, we explain current FRB detections and offer a first
 glimpse of what the true population must be.
}


\maketitle

\section{Introduction}
Fast Radio Bursts (FRBs) are bright, brief, and baffling radio transients. Since their discovery at the Parkes telescope
\citep{lorimer2007,thornton}, an array of other surveys have also detected FRBs \citep[e.g.][]{spitler2014, gbt, utmost, askap_fly,petroff_review}.
The large majority of these appear as one-off bursts, despite extensive dedicated programs of several hunderds of hours \citep[e.g.][]{2015MNRAS.454..457P,askap_nature}.
Some FRB sources have, however, been found to repeat \citep{r1,r2}.
The observed dispersion measure (DM) excess beyond the Galactic contribution puts all FRBs
at extragalactic distances, which indeed is one of their defining features.
Localized FRBs confirm this theory, showing them to originate from host galaxies other than our own at gigaparsec distances \citep{tendulkar,askap_localization,dsa_localization}.
Such FRBs allow us to start mapping out the relationship between the distance and
the dispersion measure contribution from traversing the intergalactic medium.
As a result, FRBs have been hailed as possible cosmological probes, that can in principle inform us about the intergalactic medium
\citep{macquart2013}, baryonic content \citep{mcquinn2014} or large scale structure in the universe
\citep{masui2015a}. Yet in practice, to infer the characteristics of our Universe, we need to understand e.g. the
dispersion measure contributions of the source themselves: we need to known the volumetric rate and properties of the intrinsic population.

The first ten years of the field yielded only a handful of FRB detections\footnote{For a full list of published FRBs see the FRB Catalogue: www.frbcat.org \citep{frbcat}}. Without stringent observational constraints,
no consensus on the origin of FRBs could emerge. As such, a large number of theories on the origin of FRBs have been
presented \citep[see][]{frbtheories} with suggestions ranging from young pulsars \citep[e.g.][]{connor2016supernova} to Active
Galactic Nuclei  \citep[AGNs; e.g.][]{agn}. The advent of all-sky surveys such as CHIME \citep{chimeoverview}, and of
surveys with a high spatial and fluence precision such as ASKAP \citep{askap_nature} and Apertif \citep{apertif} will drastically change
this field.
Due to their high detection rates and improved localizations the observable FRB population will be mapped much more
thoroughly. This presents the next challenge: to determine the nature of FRBs from this observed population.

With high FRB detection rates on the horizon, it is essential we understand what the detected FRBs
represent.
Directly taking the observed properties of an FRB population as representative of the underlying source class will
often be wrong.
Indeed, a variety of selection effects will prohibit a direct match, whether due to e.g.~telescope sensitivity, wavelength range, search parameters or even time resolution. Such seemingly obvious selection effects tells us that similar selection effects
must, potentially more subtly, be at play for many other FRB traits.
It is therefore essential that the mix of intrinsic FRB properties, propagation effects and selection effects are understood.

Population synthesis is a method through which the details of an intrinsic source population can be probed.
Population synthesis provides statistical insights into the parent population, and is helpful when the number of observed sources is small,
and where observational biases cannot easily be corrected for analytically.
This method is especially powerful when the underlying class is much larger and potentially more diverse than the
population that is observed.
In practice, population synthesis thus consists of three components: modelling a population, applying selection effects by modelling a survey and comparing the simulated results to real detections. This process is then repeated by adapting the modelled population or modelled survey until the results are in good agreement with each other. Each iteration in synthesising populations or modelling selection effects allows an increasingly accurate model of the underlying population to be built. In this way population synthesis not only provides insight into an intrinsic source population, but also into the often complex convolution of selection effects.

This method has previously been applied successfully to a variety of astronomical phenomena,
such as  pulsars \citep{tm77},
gamma ray bursts \citep{GRBs},
and stellar evolution \citep{2018arXiv180806883I}.
Like the FRBs under consideration in this work, pulsars are time domain sources, and many of the selection effects are identical.
\citet{go70} started with fewer pulsars, 41, than there are FRBs now, and as period derivatives had not yet been measured for most,
very little was known about these pulsars. When new surveys had increased the detected sample ten fold,
\citet{lmt85} could estimate birth rates and \citet{bwhv92} determined magnetic field longevity.
Using the modern sample of over 2,000
pulsars, statistical studies of e.g.,
radio beaming fractions \citep{lv04},
birth locations \citep{2006ApJ...643..332F}, and
radio luminosities \citep{2014ApJ...784...59S}
have improved our understanding of the pulsar population.
Such parent populations can be used to optimise the strategies for such pulsar surveys
 as using LOFAR \citep{ls10} and the Square Kilometre Array
\citep[SKA;][]{2009A&A...493.1161S},
and predict the outcomes to within a factor of a few \citep[cf.][]{2019A&A...626A.104S}.

Unfortunately next to none of the synthesis codes that produced the work mentioned above were made public. And thus, for
example, an argument over two versus one pulsar birth populations (\citealt{no90} versus \citealt{bwhv92}) was at least partly
fueled by incomplete understanding of the used codes, which were  both proprietary and closed.
The synthesis work by \citet{2009A&A...493.1161S}, and by \cite{2006MNRAS.372..777L}, however, \emph{were} reproducible,
because they were based on
\texttt{PSRPOP} \citep{2011ascl.soft07019L} and
\texttt{PsrPopPy} \citep{psrpoppy, 2015ascl.soft01006B}.

Prior efforts at FRB population synthesis have mostly been directed towards dedicated surveys.
A number looked primarily into FRB volumetric densities \citep[e.g.][]{caleb2016, fialkov2018, niino2018, Bhattacharya2019}, with others focused on the origin of the excess dispersion measure \citep{walker2018}, on spectral indices \citep{chawla2017}, on brightness distributions \citep{Oppermann16, vedantham2016, jp2, jp1} and on repeat fractions \citep{caleb2019}. Despite a large variety of FRB population synthesis models, the underlying code is not always provided or easily adaptable.

It is important FRB detections are reported with a full understanding of underlying selection effects and by extension their relation to the intrinsic FRB population. An open platform for FRB population synthesis can facilitate that, which is why we have developed \frbpoppy (Fast Radio Burst POPulation synthesis in PYthon). This open source software package aims to be modular and easy-to-use, allowing survey teams to understand implications of new detections. \frbpoppy can help in the study of FRB population features and in predicting future results, just as pulsar population synthesis did for the pulsar community.

In this paper we aim to determine what the real FRB parent population must look like,
and we present the first version of \frbpoppy (\texttt{v1.0.0}), accessible on
Github\footnote{https://github.com/davidgardenier/frbpoppy}.
We start the paper by describing \frbpoppy's simulation process, before demonstrating some applications of the code in
latter half of the paper. Accordingly, Sect.~\ref{sec:generate} describes how an intrinsic FRB population is simulated,
Sect.~\ref{sec:observe} describes how a survey is simulated, Sect.~\ref{sec:real} describes how real detections are
used and Sect.~\ref{sec:comp} details how we compare simulated and real FRB populations.
In Sect.~\ref{sec:results} we describe results, and in Sect.~\ref{sec:discussion} we discuss
how a simple, local population of standard candles cannot describe current observations.
A cosmological population, with a specific distribution of pulse widths, spectral indexes, and luminosities is required
to reproduce the observed FRB sky. The paper is rounded off with a conclusion in Sect.~\ref{sec:conclusion} and additional information in appendix~\ref{sec:appendix:beam}.

\section{Generating an FRB population}
\label{sec:generate}
The main goal in population synthesis is to infer the properties of the real, underlying parent population, through a simulated population.

Following conventions in pulsar population synthesis \citep[e.g.][]{bwhv92}, we aim to keep a clear distinction between these real and simulated FRB populations. Both real and simulated experiments deal with two sets of distributions: The population’s intrinsic physical properties, including their luminosity function and redshift distribution, as well as their observed properties, for example the brightness and DM distributions. We use the terms ``underlying’’, ``parent’’, and ``progenitor’’ interchangeably with the former, and we use the phrase ``surveyed’’ or ``detected'' synonymously with ``observed’’. We refer to FRBs generated/observed in the \frbpoppy framework as \textit{simulated} and actual FRBs as \textit{real}.

Our method  consists of three parts:
modelling an intrinsic population, applying selection effects and
comparing the simulated population to real detections.
Out of these three components,
it is the modelling of an
intrinsic FRB population that allows the underlying physics to be probed.
This we do by first formulating a hypothesis on what the parent population is, and how it behaves.
We subsequently translate this to the
parameters available in \frbpoppy, listed in Table~\ref{tab:populations}.
These can be adjusted to simulate e.g. different source-class densities and emission characteristics, or propagation effects.
By doing this for various models, running the population synthesis separately on each,
and comparing the outcome (cf.~Sect.~\ref{sec:discussion}),
we can learn which underlying population best describes the observed FRB sky.
In this paper we compare four models.
The adopted values for each are listed in Table~\ref{tab:populations}.
Using these, we aim to answer questions such as: does the host dispersion measure have a measurable influence on the
population our telescopes detect? Can a model employing standard candles reproduce the observed fluence distributions?
Subsequent sections describe each of the model aspects.

\begin{table*}
 \caption{Overview of parameters required to generate an initial cosmic FRB population. The four population setups given in this table are labeled with the terms \pop{Default}, \pop{Simple}, \pop{Complex} and \pop{Standard Candles}, each describing the defining characteristic of the population.
  \label{tab:populations}
 }
 \centering
 \begin{tabular}{c c c c c c}
  \hline\hline
  Parameters                      & Units     & \pop{Default}                    & \pop{Simple}      & \pop{Complex}     & \pop{Standard Candles} \\
  \hline
  $\text{n}_{\rm model}$          &           & SFR                              & vol$_{\rm co}$    & vol$_{\rm co}$    & SFR                    \\
  $\text{H}_{0}$                  & km/s/Mpc  & 67.74                            & 67.74             & 67.74             & 67.74                  \\
  $\Omega_{\rm m}$                &           & 0.3089                           & 0.3089            & 0.3089            & 0.3089                 \\
  $\Omega_{\Lambda}$    &           & 0.6911                           & 0.6911            & 0.6911            & 0.6911                 \\
  $\text{DM}_{\rm host,\ model}$  &           & normal                           & normal            & normal            & normal                 \\
  $\text{DM}_{\rm host,\ \mu}$    & pc/cm$^3$ & 100                              & 0                 & 100               & 100                    \\
  $\text{DM}_{\rm host,\ \sigma}$ & pc/cm$^3$ & 200                              & 0                 & 200               & 0                      \\
  $\text{DM}_{\rm igm,\ index}$   & pc/cm$^3$ & 1000                             & 0                 & 1000              & 1000                   \\
  $\text{DM}_{\rm igm,\ \sigma}$  & pc/cm$^3$ & $0.2\text{DM}_{\rm igm,\ index}$ & 0                 & $200z$            & $200z$                 \\
  $\text{DM}_{\rm mw,\ model}$    &           & NE2001                           & zero              & NE2001            & NE2001                 \\
  $\nu_{\rm emission, range}$     & MHz       & $10^6$-$10^9$                      & $10^6$-$10^9$       & $10^6$-$10^9$       & $10^6$-$10^9$            \\
  $\text{L}_{\rm bol,\ range}$    & ergs/s    & $10^{39}$-$10^{45}$                & $10^{38}$-$10^{38}$ & $10^{39}$-$10^{45}$ & $10^{36}$-$10^{36}$      \\
  $\text{L}_{\rm bol,\ index}$    &           & 0                                & 0                 & 0                 & 0                      \\
  $\alpha_{\rm in}$               &           & $-1.5$                             & $-1.5$              & $-1.5$              & $-1.5$                   \\
  $w_{\rm int,\ model}$           &           & Lognormal                        & Uniform           & Lognormal         & Uniform                \\
  $w_{\rm int,\ range}$           & ms        & 0.1-10                           & 10-10             & -                 & 1-1                    \\
  $w_{\rm int,\ \mu}$             & ms        & 0.1                              & -                 & 0.1               & -                      \\
  $w_{\rm int,\ \sigma}$          & ms        & 0.5                              & -                 & 0.7               & -                      \\
  $\gamma_{\mu}$                  &           & $-1.4$                             & 0                 & $-1.4$              & 0                      \\
  $\gamma_{\sigma}$               &           & 1                                & 0                 & 1                 & 0                      \\
  $z_{\rm max}$                   &           & 2.5                              & 0.01              & 2.5               & 2.5                    \\
  $\text{n}_{\rm gen}$            &           & -                                & $10^8$            & $10^8$            & $10^8$                 \\
  \hline
 \end{tabular}
\end{table*}

\subsection{Number density}
\label{sec:generate:density}
What volumetric rate of FRB progenitors is needed to reproduce the observed sample?
To establish the underlying number density of FRB sources,
we model a number of population characteristics.
In the work presented here, we limit ourselves to one-off FRBs and
leave the treatment of repeating FRBs to the near future.
All FRB setups generate sources isotropically distributed on the sky;
with individual distances being set by the following source number density models:\\

\emph{Constant}
FRBs have a constant number density per comoving volume element such that
\begin{equation}
    \rho_{\rm FRB}(z) = C
\end{equation}
with $\rho_{\rm FRB}(z)$ the constant number density of FRBs such that there is no redshift dependence.
Given $\rho_{\rm FRB}(z) = dN/dV_{\rm co}$ with the differential number of FRBs $dN$ in a comoving volume element $dV_{\rm co}$, $dN = \rho_{\rm FRB}(z) \cdot dV_{\rm co} = C \cdot dV_{\rm co}$ and so $dN \propto dV_{\rm co}$. We can therefore simulate a constant number density distribution by uniformly sampling a comoving volume $V_{\rm co}$ space. In \frbpoppy we convert a given maximum redshift to the corresponding maximum comoving volume such that this space can be sampled using:
\begin{equation}
 V_{\rm co,\ FRB} = V_{\rm co,\ max} \cdot U(0, 1)
 \label{eq:vol_co}
\end{equation}
with the comoving volume of an FRB $V_{\rm co,\ FRB}$, the maximum comoving volume $V_{\rm co,\ max}$ and a random number from a uniform distribution with $U \in [0,1]$. Conversions to e.g.~luminosity distance and redshift are based on \citet{wright2006} using cosmological parameters from \citet{planck}, of which the latter can be found in Table~\ref{tab:populations}. \\

\emph{Star Formation Rate (SFR)}
The FRBs number density is proportional to the comoving star formation rate. Using \citet{madau2014}, FRBs are distributed according to
\begin{equation}
 \rho_{\rm FRB}(z) \propto \frac{(1+z)^{2.7}}{1+[(1+z)/2.9]^{5.6}}.
 \label{eq:sfr}
\end{equation}\\
with $\rho_{\rm FRB}(z)$ the comoving number density of FRBs at a given redshift $z$. We sample this distribution by numerical constructing a cumulative distribution function (CDF) of Eq.~\ref{eq:sfr} over redshift. Uniformly sampling this CDF provides the corresponding redshift distribution which can then be converted to any other required cosmological distances.\\

\emph{Stellar Mass Density (SMD)}
FRBs follow the relationship between redshift and cosmic stellar mass density as given by \citet{madau2014}, using
\begin{equation}
 \rho_{FRB}(z) \propto \int_{z}^\infty \frac{(1+z')^{1.7}}{1+[(1+z')/2.9]^{5.6}} \frac{{\rm d}z'}{H(z')}
 \label{eq:smd}
\end{equation}
with $\rho_{FRB}(z)$ the number density at redshift $z$ and $H(z')$ the Hubble parameter in a flat cosmology such that the spatial curvature density parameter $\Omega_k$ is zero. $H(z')$ can then be further defined as
\begin{equation}
 H(z')=H_0 \sqrt{\Omega_m(1+z')^3 + \Omega_\Lambda}
\end{equation}
with the Hubble parameter $H_0$, the matter density parameter $\Omega_m$ and the dark energy density parameter $\Omega_\lambda$ \citep{madau2014}. We simulate the SMD in a manner similar to the SFR: we first construct a CDF over redshift for Eq.~\ref{eq:smd} which we then uniformly sample to obtain a redshift distribution.\\

\label{sec:alpha}
\emph{Power law}
While a constant number density per comoving volume may work in many cases, the ability to vary this density can be
helpful. For example, modelling a relative overabundance of local FRBs can prove interesting.
To this end, we also model various density-distance relations with
\begin{equation}
 V_{\rm co,\ FRB} = V_{\rm co, max} \cdot U(0, 1) ^ \beta
 \label{eq:vol_co_pl}
\end{equation}
following Eq.~\ref{eq:vol_co} in setting $V_{\rm co}$, but instead scaling the uniform sampling $U(0, 1)$ by $\beta$. This exponent $\beta$ allows for instance for relatively more local sources and less distant sources to be generated. Once combined with the luminosity function and the instrument response, this number density relation to distance (and hence, fluence) will determine the observed brightness distribution of FRBs.

Rather than using $\beta$ as input, out of convenience a different expression can be used:
\begin{equation}
 \beta = - \frac{3}{2 \alpha_{\rm in}}
 \label{eq:beta}
\end{equation}
with $\beta$ the power as given in Eq.~\ref{eq:vol_co_pl} and $\alpha_{\rm in}$ an input parameter. In a Euclidean universe, the total number of sources $N$ out to a radius $R$ scales as $N({<}R) \propto R^3$. Combined with the flux $S$ scaling as $S \propto R^{-2}$, for standard candles one can derive $N({>}S) \propto S^{-3/2}$. This exponent of the $\log N$-$\log S$ relation can also be expressed as $\alpha$, so for a Euclidean universe $\alpha$ is expected to equal $-3/2$.
However, when a power law relation is chosen in \frbpoppy, these relationships change. Instead due to the change in sampling the comoving volume, $N({<}R)^\beta \propto R^{3}$, or $N({<}R) \propto R^{3/\beta}$ leading to $N({>}S) \propto S^{-3/(2\beta)}$. Given Eq.~\ref{eq:beta}, this is equivalent to saying $N({>}S) \propto S^{\alpha_{in}}$. Eq.~\ref{eq:beta} therefore allows $\alpha_{\rm in}$ to have a value such if a Euclidean population was observed with a perfect survey, $\alpha_{\rm in}$ would equal the observed slope $\alpha$ of the $\log N$-$\log S$ relation. In different words, within FRB detection completeness in the very nearby universe, $\alpha = \alpha_{\rm in}$. Extending this to cosmological distances says that surveying any FRB population with a given $\alpha_{\rm in}$ would result in an observed $\log N$-$\log S$ slope asymptoting towards $\alpha_{\rm in}$ in the limit of the local universe. An extensive discussion of this topic can be found in \citet{jp2}. \\

\begin{figure}
 \centering
 \includegraphics{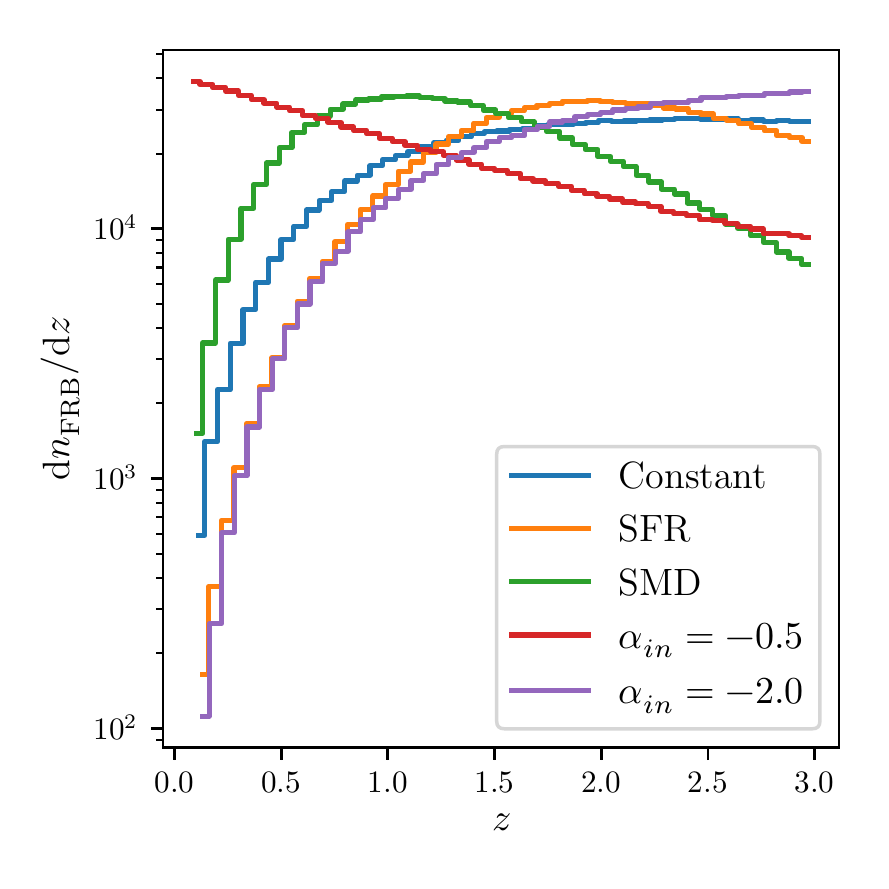}
 \caption{Comoving number density ($\rho(z)\equiv dN/dz$) as a function of redshift from a simulated distribution of $10^6$~FRBs. The FRBs follow either a constant number density per comoving volume element \citep{wright2006}, the star formation rate (SFR) \citep{madau2014},  the stellar mass density (SMD) \citep{madau2014}, or a power law in the comoving volume space with index $\alpha_{in}=-0.5$ or $\alpha_{in}=-2.0$ (see Sec.~\ref{sec:alpha}). Note that $dn_{\rm FRB}$ refers to the intrinsic number of FRBs rather than an observed number, as that would be affected by a factor of $(1 + z)^{-1}$ as well the luminosity function, spectral index etc.}
 \label{fig:num_den}%
\end{figure}

Fig.~\ref{fig:num_den} shows five populations following the models described above, with the constant number density population showing clear cosmological effects with increasing redshift. This behaviour, in which the number density flattens out at larger redshifts, is as expected due to volume running out towards larger cosmological distances. The corresponding comoving volume $V(z)$ at redshift $z$ matches those as calculated using \citet{hogg1999}.

~\\

\subsection{Dispersion measure}
The dispersion measure quantifies the relative arrival time of an FRB with respect to its highest frequency and is defined such that
\begin{equation}
    {\rm DM} = \int^{d}_{0}n_e {\rm d}l
\end{equation}
with the rest frame dispersion measure ${\rm DM}$, distance $d$, electron number density $n_e$ and differential step ${\rm d}l$ \citep{handbook}. This measure represents the column density of free electrons along the line of sight, but is mute on the location of these electrons. Most cosmology tests with FRBs require an understanding of the various contributors to the total observed dispersion measure.
We thus model the total dispersion measure as the addition of several components, to aid in assigning different certainties and models to each:
\begin{equation}
 \label{eq:dm}
 {\rm DM}_{\rm tot} = \frac{{\rm DM}_{\rm host}}{1+z} + {\rm DM}_{\rm IGM} + {\rm DM}_{\rm MW}
\end{equation}
with the total dispersion measure ${\rm DM}_{\rm tot}$, the dispersion measure contribution by the host galaxy ${\rm DM}_{\rm host}$ adapted by the redshift~$z$ \citep{tendulkar}, the contribution from the intergalactic medium ${\rm DM}_{\rm IGM}$ and finally the Milky Way component ${\rm DM}_{\rm MW}$.

\subsubsection{Dispersion measure - host}
Lacking strong constraints on the host galaxy dispersion measure, we default to \citet{thornton} in adopting a value of $100\ \text{pc}\ \text{cm}^{-3}$, and adding a Gaussian spread to this value while ensuring ${\rm DM}_{\rm IGM}
 >0\ \text{pc}\ \text{cm}^{-3}$. Using such a distribution, we can replicate observations seeming to indicate a varying dispersion measure contribution from the host galaxy and/or source \citep[e.g.][]{michilli}. A variety of models are available in \frbpoppy, allowing the ${\rm DM}_{\rm IGM}$ to more accurate represent future observations.

\subsubsection{Dispersion measure - IGM}
Modeling the free electron density in the intergalactic medium is a challenging task, whether in disentangling contributions from the Milky Way or host, or even in obtaining observations capable of probing this intervening matter. While often an approximation of $\text{DM}_{\text{IGM}} \approx 1200z$ is used for the intergalactic medium contribution to the total dispersion measure \citep{ioka2003,inoue2004}, recent research seems to be tending towards values in the range of 800-1000~pc/cm$^{-3}$ \citep[e.g.][]{zhang_dm,keane2018,pol2019}, or non-linear relationships such as given in \citet{batten2019}. In this paper, the value of $\text{DM}_{\text{IGM}}$ for an FRB at redshift $z$ is drawn from a Gaussian distribution $\mathcal{N}(1000z,200z)$ with $\mathcal{N}(\mu, \sigma)$  denoting the values for the mean $\mu$ and a standard deviation $\sigma$. In this way, a scatter around a linear relationship between $\text{DM}_{\text{IGM}}$ and $z$ is introduced. This method can be updated as new information becomes available \citep[e.g.][]{bannister2019, ravi2019tomography}.

\subsubsection{Dispersion measure - Milky Way}
With over 50 years of pulsar observations \citep{Hewish1968},
the Galactic dispersion measure has better constraints
than that of the intergalactic medium. For the current work, we use \citet{ne2001}.
Developed as a tool to estimate pulsar dispersion measures in the Milky Way, it is a familiar model to those working in the field, despite some of its distance measurements being older than those in e.g., \citet{ymw}.
We use the dispersion measure values taken in each direction queried at a distance of 100~kpc to retrieve the maximum Galactic dispersion measure. This distance also surpasses the maximum radial extent of the thick disk of 20~kpc in every direction \citep{ne2001-2}.
Other models can of course be added to \frbpoppy by those interested.

\subsection{Luminosity}
Determining the correct intrinsic FRB luminosity distribution may tell us how FRBs radiate.
A number of radiation models have been suggested \citep[e.g.][]{katz2014,romero2016,lu2018,metzger2019,beloborodov2019}.
Without
observational constraints on the intrinsic emission mechanism of FRBs, sources in \frbpoppy are assumed to be radiating isotropically.
Luminosities are generated following a power law distribution, with options to set the
index (L$_{\rm bol,\ index}$),
and the  minimum and maximum value (L$_{\rm bol,\ range}$).
While work by e.g. \citet{caleb2016} also adopt power law functions, recent work by \citet{luo2018} and \citet{fialkov2018} indicate a Schechter luminosity function might provide a more accurate description. While in this initial version of \frbpoppy we only include a power law model, other distributions such as a Schechter luminosity function or a broken power law could be implemented in future iterations.

\subsection{Spectral index}
To further understand the FRB emission process,
we aim to learn whether they emit over a wide spectrum,
and at which frequencies they are brightest.
In \frbpoppy, as in \texttt{psrpoppy}, we thus allow the intrinsic spectral indices for individual FRBs to be drawn from a Gaussian
distribution for which the mean and standard deviation can be set.
We define the spectral index $\gamma$ such that
\begin{equation}
 E_{\nu} = k \nu^{\gamma}
\end{equation}
with the energy $E_{\nu}$ at the rest-frame frequency $\nu'$ \citep{lorimer2013}.
As the intrinsic spectral index of the FRB population has proven difficult to determine \citep[e.g.][]{spitler2014, scholz2016},
we draw $\gamma$ from a Gaussian distribution centred around -1.4 with a standard deviation of 1,
as in \citet{bates2013}. This replicates observations of the Galactic pulsar population. Recent work by \citet{macquart2019spectral} favours similar values for the FRB population.

\subsection{Pulse width}
Determining the intrinsic FRB pulse widths can elucidate some very specific traits
of the source environment, such the size of the emitting region, or the beaming fraction for a rotating source.
As the observed  FRB pulse width detections cluster around millisecond timescales,
we use as input one of two models: \\

\emph{Uniform} pulse width values are randomly chosen between a given lower and higher millisecond timescale. \\

\emph{Lognormal} In order to replicate the distribution of pulse widths observed in pulsars, or indeed
repeater pulses we default to drawing pulse widths from a log-normal distribution. The probability density function such that a variable $x$ is considered to have a log-normal distribution can be expressed as
\begin{equation}
    p(x) = {\frac {1}{x}} \cdot {\frac {1}{\sigma {\sqrt {2\pi \,}}}}\exp \left(-{\frac {(\ln x-\mu )^{2}}{2\sigma ^{2}}}\right)
\end{equation}
for the variable $x$, standard deviation $\sigma$ and mean $\mu$ \citep{johnson1994continuous}. \frbpoppy provides options to adapt the mean and standard deviation of this distribution, which can be adjusted to replicate broad or narrow pulse widths.

\subsection{Number of sources}
Internally, the simulated FRB population will be formed by a certain total number of sources ($\text{n}_{\rm gen}$). The value of this parameter will depend on the resolution sought in the resulting population, while taking a wide range of selection effects into account. Based on results from the high-latitude HTRU survey, \citet{thornton} measured an FRB rate of $1.0^{+0.6}_{-0.5} \times 10^4\ \text{sky}^{-1}\text{day}^{-1}$ above a $3\ \text{Jy\ ms}$ threshold.
Subsequent detections updated the rate to  $6^{+4}_{-3}\ \times 10^3\ \text{sky}^{-1}\text{day}^{-1}$ \citep{champion}, and taking completeness into account \citet{keane2015completeness} measured a rate of $2500\ \text{sky}^{-1}\text{day}^{-1}$ above a $1.4$-GHz fluence of $2\ \text{Jy\ ms}$.
Therefore, unless seeking to use a \survey{perfect} survey, i.e.~a survey in which all FRBs are detected, cosmic FRB populations should be generated with ${>}10^4$ FRBs to ensure sufficient simulated detections. Population and survey parameter choices have a strong influence on this number, and as such this value is given solely as a very rough indication.

\subsection{Number of days}
Setting the number of days over which a population of FRBs is emitted ($\text{n}_{\rm days}$) provides a way to set a volumetric rate. Within this paper, all detection rates are scaled relative to each other, and accordingly the number of days is set to one. Users can however use this parameter, coupled with the number of survey days to get a simulated absolute detection rate. Matching this to a real detection rate allows the volumetric rate of FRBs to be probed.

\section{Observing an FRB population}
\label{sec:observe}
The observed FRB population will always differ from the intrinsic one  - the former involves a number of selection effects layered
on top of the intrinsic FRB population \citep{connor2019}.
The following section describes how we construct virtual surveys, each with
different e.g. celestial selection effects, and hardware constraints.

\subsection{Surveys}
The telescope with which a survey is conducted can cause a large variety of
selection effects. For example, surveys
are biased against detecting both narrow pulses and highly-dispersed pulses, as
the finite time and frequency resolution of the instruments results in
deleterious smearing effects \citep{connor2019}. The strength  of such
hardware selection effects however can vary per survey. These very same
selection effects have been long known to be highly important  for pulsar
surveys \citep[e.g.,][]{tm77}.

In Table~\ref{tab:surveys} we present an overview of the survey parameters adopted within \frbpoppy.
Using these parameters a survey model can be constructed.
From these, we infer the resultant selection effects, to model the expected survey rates and parameter distributions.
While the  Table~\ref{tab:surveys} values are sufficient to reproduce the results found in the current work,
additional
surveys are already included in \frbpoppy, and new ones are easy to implement.
CHIME, for instance, already detects FRBs at a very high rate, but it
is not included in this work because we are not yet sufficiently confident modelling its system parameters.
It is also the only survey with detections below 700~MHz.
Still, an early version of this survey model is included in \frbpoppy, and subsequent research will cover CHIME detections.

\begin{sidewaystable}
  \centering
 \caption{An overview of survey parameters used in this paper. Parameters include the survey degradation factor $\beta$, telescope gain $G$, sampling time $t_{\textrm{samp}}$, receiver temperature $T_{\textrm{rec}}$, central frequency $\nu_{\textrm{c}}$, bandwidth BW, channel bandwidth BW$_{\textrm{ch}}$, number of polarisations $n_{\textrm{pol}}$, Field-of-View FoV, minimum signal to noise ratio S/N and then the minimum to maximum right ascension $\alpha$, declination $\delta$, Galactic longitude $l$ and Galactic latitude $b$.
  While the majority of parameter are drawn from the references as given below the table, a number of parameters have
  been calculated as an average between given values, estimated or acquired through private communication. These are
  denoted in grey.
 }
 \label{tab:surveys}
 \centering
\begin{tabular}{cccccccccccccccc}
  \hline
  \hline
 Parameter            & $\beta$   & G            & $t_{\textrm{samp}}$ & $T_{\textrm{rec}}$ & $\nu_{\textrm{c}}$ & BW        & BW$_{\textrm{ch}}$ & $n_{\textrm{pol}}$ & FoV         & S/N              & $\alpha$       & $\delta$             & $l$                 & $b$               & References \\
 Units                &           & K/Jy         & ms                  & K                  & MHz                & MHz       & MHz                &                    & deg$^2$     &                  & \degr          & \degr                & \degr               & \degr             &            \\
 \hline
 \survey{apertif}     & \est{1.2} & \est{1.1}    & \est{0.04096}       & {70}               & {1370}             & {300}     & {0.19531}          & \est{2}            & {8.7}       & \est{8}          & \est{0 -- 360} & \est{$-$37.1 -- 90}  & \est{$-$180 -- 180} & \est{$-$90 -- 90} & 1          \\
 \survey{askap-fly}   & \est{1.2} & \est{0.035}  & {1.265}             & \est{70}           & {1320}             & {336}     & {1}                & {2}                & {160}   & \est{8}          & \est{0 -- 360} & \est{$-$90 -- 40}    & \est{$-$180 -- 180} & \est{$-$90 -- 90} & 2          \\
 \survey{askap-incoh} & \est{1.2} & \est{0.1}    & {1.265}             & \est{200}          & {1320}             & {336}     & {1}                & {2}                & {20}        & \est{8}          & \est{0 -- 360} & \est{$-$90 -- 40}    & \est{$-$180 -- 180} & \est{$-$90 -- 90} & 2          \\
 \survey{gbt}         & \est{1.2} & {2}          & {1.024}             & {1.16}             & {800}              & {200}     & {0.05}             & {2}                & {0.016}     & {8}              & \est{0 -- 360} & \est{$-$51.57 -- 90} & \est{$-$180 -- 180} & \est{$-$90 -- 90} & 3          \\
 \survey{htru}        & \est{1.2} & {0.69}       & {0.064}             & \est{28}           & {1352}             & {340}     & {0.390625}         & {2}                & {0.56}      & {8}              & \est{0 -- 360} & \est{$-$90 -- 90}    & {$-$120 -- 30}      & {$-$15 -- 15}     & 4          \\
 \survey{palfa}       & \est{1.2} & {8.2}        & {0.0655}            & \est{26}           & {1375}             & {322}     & {0.390625}         & {2}                & {0.022}     & {8}              & \est{0 -- 360} & {$-$5. -- 35}        & {30 -- 78}          & {$-$5 -- 5}       & 5          \\
 \survey{parkes}      & \est{1.2} & \est{0.69}   & \est{0.064}         & \est{28}           & \est{1352}         & \est{340} & \est{0.390625}     & \est{2}            & \est{0.56}  & \est{8}          & \est{0 -- 360} & \est{$-$90 -- 47}    & \est{$-$180 -- 180} & \est{$-$90 -- 90} & 4          \\
 \survey{perfect}     & \est{1.2} & \est{100000} & \est{0.001}         & \est{0.001}        & \est{1000}         & \est{800} & \est{0.001}        & \est{2}            & \est{41253} & \est{$10^{-16}$} & \est{0 -- 360} & \est{$-$90 -- 90}    & \est{$-$180 -- 180} & \est{$-$90 -- 90} & \\
 \survey{utmost}      & \est{1.2} & {3.6} & {0.65536}         & {400}        & {843}         & {16} & {0.78125}        & {1}            & {7.80} & {10} & \est{0 -- 360} & {$-$90 -- 18}    & \est{$-$180 -- 180} & \est{$-$90 -- 90} & 6 \\
 \hline
\end{tabular}

 \vspace{1ex}
 \raggedright \emph{1}~\citet{oosterloo, apertif} \emph{2}~\citet{chippendale, askap_fly, askap_nature} \emph{3}~\citet{gbt} \emph{4}~\citet{htru} \emph{5}~\citet{cordes_palfa, lazarus, patel} \emph{6}~\citet{caleb2016utmost, 2017PASA...34...45B}

\end{sidewaystable}

\subsection{Pulse width}
A variety of effects modify the FRB pulses as they travel through space, and are detected on Earth.

The first effect is purely cosmological.
Depending on the method used to populate the simulated FRB event space, a comoving distance might need to be calculated from a redshift distribution, or the inverse. With both of these taking place over large distances, cosmology must be taken into account when calculating parameter values upon arrival at Earth, rather than simply taking the initial value. The pulse width of an FRB arriving at Earth is then
\begin{equation}
 w_{\rm arr} = (1+z)w_{\rm int}
\end{equation}
where the intrinsic pulse width $w_{\rm int}$  at redshift $z$ has been dilated to the pulse width as it arrives at Earth,  $w_{\rm arr}$.

The second effect, in principle, is the increase of the observed pulse width due to multi-path scattering.
In  \frbpoppy the parameter $t_{\rm scat}$ allows scattering timescales to be included in calculating the effective
pulse width.
The adaptation of \citet{bhat2004} to FRBs from \citet{lorimer2013} is included in \frbpoppy, being
\begin{equation}
 \log t_{\rm scat} = -9.5 + 0.154 (\log {\rm DM_{\rm tot}}) +  1.07 (\log {\rm DM_{\rm tot}})^2 -3.86 \log \nu_{\rm c}
\end{equation}
with the scattering timescale $t_{\rm scat}$, the total dispersion measure ${\rm DM_{\rm tot}}$, and the central survey frequency in GHz $\nu_{\rm c}$. A Gaussian scatter is subsequently applied such that
\begin{equation}
  t_{\rm scat} = 10^{\mathcal{N}(\log t_{\rm scat},\ 0.8)}
\end{equation}
with the scattering timescale $t_{\rm scat}$ and a Gaussian function $\mathcal{N}(\mu, \sigma)$ with the mean $\mu$ and standard deviation $\sigma$.
The current FRB population appears to be underscattered relative to Galactic pulsars \citep[see e.g.][]{ravi2019prevalence}. Many FRB profiles show the presence of scattering; however, no consistent scattering relation has yet been established and a larger future population may be needed.
Due to our incomplete understanding regarding the scattering properties of FRBs \citep[see e.g.][]{cordes2016,xu2016}, we set the scattering timescale as a default to zero.

Thirdly we take into account the effects of intra-channel dispersion smearing $t_{\rm DM}$,
and the sampling time $t_{\rm samp}$.
Starting with the dispersion smearing, $t_{\rm DM}$ can be calculated following
\begin{equation}
 t_{\rm DM} = 8.297616\cdot 10^6 \cdot ({\nu_2-\nu_1}) \cdot {\rm DM_{\rm tot}} \cdot \nu_{\rm c}^{-3}
\end{equation}
with the dispersion smearing $t_{\rm DM}$ in ms, the lower and upper frequency of a survey channel respectively $\nu_1$ and $\nu_2$, and the central frequency thereof $\nu_{\rm c}$, all in MHz \citep{cordes2003}, and the total dispersion measure ${\rm DM_{\rm tot}}$ as given in equation~\ref{eq:dm}.

The final term is the sampling timescale $t_{\rm samp}$. This is provided as input per survey and can be found in Table ~\ref{tab:surveys}.

Together these contribute to the observed pulse with $w_{\rm eff}$ are added as
\begin{equation}
 \label{eq:w_eff}
 w_{\rm eff} = \sqrt{w_{\rm arr}^2 + t_{\rm scat}^2 + t_{\rm DM}^2 + t_{\rm samp}^2}
\end{equation}
and it is that pulse width that is used in determining whether the FRB is detected \citep{handbook}.

\subsection{Detection}
The brightness detection threshold of an FRB can be determined by the radiometer equation for a single pulse:
\begin{equation}
 {\rm S/N} = \frac{\bar{S}_{\rm peak} G}{ \beta T_{\rm sys}} \sqrt{n_{\rm pol} ({\nu_2-\nu_1}) w_{\rm arr}}
 \label{eq:S/N}
\end{equation}
with the peak flux density $\bar{S}_{\rm peak}$, the gain $G$, degradation factor $\beta$, total system temperature $T_{\rm
   sys}$, the number of polarisations $n_{\rm pol}$, the boundary frequencies of a survey $\nu_{1,2}$ and the pulse width
at Earth $w_{\rm arr}$ \citep{handbook, connor2019}. As the system temperature
\begin{equation}
 T_{\rm sys} = T_{\rm rec} + T_{\rm sky}
\end{equation}
with the receiver temperature $T_{\rm rec}$ and sky temperature $T_{\rm sky}$ \citep{handbook}, $T_{\rm rec}$ joins $G$, $\beta$, $n_{\rm pol}$ and $\nu_{1,2}$ as survey dependent parameters, and can be found in Table~\ref{tab:surveys}.
We take $T_{\rm sky}$  to be dominated by synchrotron radiation and  scale it as
\begin{equation}
 T_{\rm sky} = T_{\rm 408\ MHz} \left(\frac{\nu_{\rm c}}{408\ {\rm MHz}}\right)^{-2.6}
\end{equation}
with the directional dependent values from the 408~MHz sky survey $T_{\rm 408\ MHz}$ and the central frequency $\nu_{\rm c}$ \citep{haslam408}. Returning to Eq.~\ref{eq:S/N}, and taking cosmology into account, $\bar{S}_{\rm peak}$ can be calculated with
\begin{equation}
 \bar{S}_{\rm peak} = \frac{L_{\rm bol} \; (1+z)^{\gamma-1}}
 {4 \pi D(z)^2 (\nu'^{\gamma+1}_{\rm high}-\nu'^{\gamma+1}_{\rm low})}
 \left(
 \frac{\nu_2^{\gamma+1}-\nu_1^{\gamma+1}}{\nu_2-\nu_1}
 \right)
 \left(
 \frac{w_{\rm arr}}{w_{\rm eff}}
 \right)
\end{equation}
with the luminosity $L_{\rm bol}$, the redshift $z$, the comoving distance $D(z)$,
and spectral index $\gamma$ \citep{lorimer2013, connor2019}.
The luminosity refers to an isotropic equivalent bolometric luminosity in the radio,
where the frequency range is defined by $\nu_{\rm low,high}$. This is because we do not include
beaming effects, and we do not attempt to model emission outside of $\nu_{\rm low}$ and $\nu_{\rm high}$.
We set the boundary emission frequencies of an FRB source $\nu_{\rm low, high}$,
to 10\,MHz and 10\,GHz as a default.
The pulse width at Earth $w_{\rm arr}$ and effective pulse width $w_{\rm eff}$ \citep{lorimer2013, connor2019} are used to take into account
the degradation of the peak flux due pulse broadening;
in effect raising the detection threshold.

The equations as given above allow a brightness threshold for an FRB detection to be set, but do not automatically equate to a detection. To that end, the S/N of each FRB must first be convolved with a beam pattern.

\subsection{Beam patterns}
A number of modelled beam patterns are available in \frbpoppy. Given the scaled angular distance on the sky from the beam centre $r \in [0,1]$, the
following beam models describe the relative sensitivity pattern $I(r)$. The link between beam patterns and observing frequency is modelled in \frbpoppy via the Field-of-View parameter as given in Table~\ref{tab:surveys}, which is presumed to be valid for the central frequency of a survey.\\

\emph{Perfect}
A perfect intensity profile, i.e.~no beam pattern, for testing, and for comparing realistic beam patterns against.
\begin{equation}
 I(r) = 1
\end{equation}

\emph{Airy}
The beam pattern of a single-dish, single-pixel radio telescope can be best described with an Airy disk, for which a simple representation can be made with
\begin{equation}
 I(r) = 4 \left(\frac{J_1\left(k \sin N(r) \right)}{k \sin N(r)}\right)^2.
\end{equation}
Derivations for the equations of the scaling factor $k$ and radial offset $N(r)$ can be found in appendix~\ref{sec:appendix:beam}. Both provide scaling factors for the Airy disk.\\

\emph{Gaussian}
An additional option is to model the intensity profile as a Gaussian beam:
\begin{equation}
 I(r) = e^{- r^2 M^2 \ln(2)}
\end{equation}
Here too, the derivation of the scaling factor $M$ can be found in appendix~\ref{sec:appendix:beam}, relating to the maximum offset. $r$ remains a normalised radial offset from the beam centre to the maximum available offset, being drawn from a uniform distribution such that $r \in [0,1]$.

\begin{figure}
 \centering
 \includegraphics[width=\linewidth]{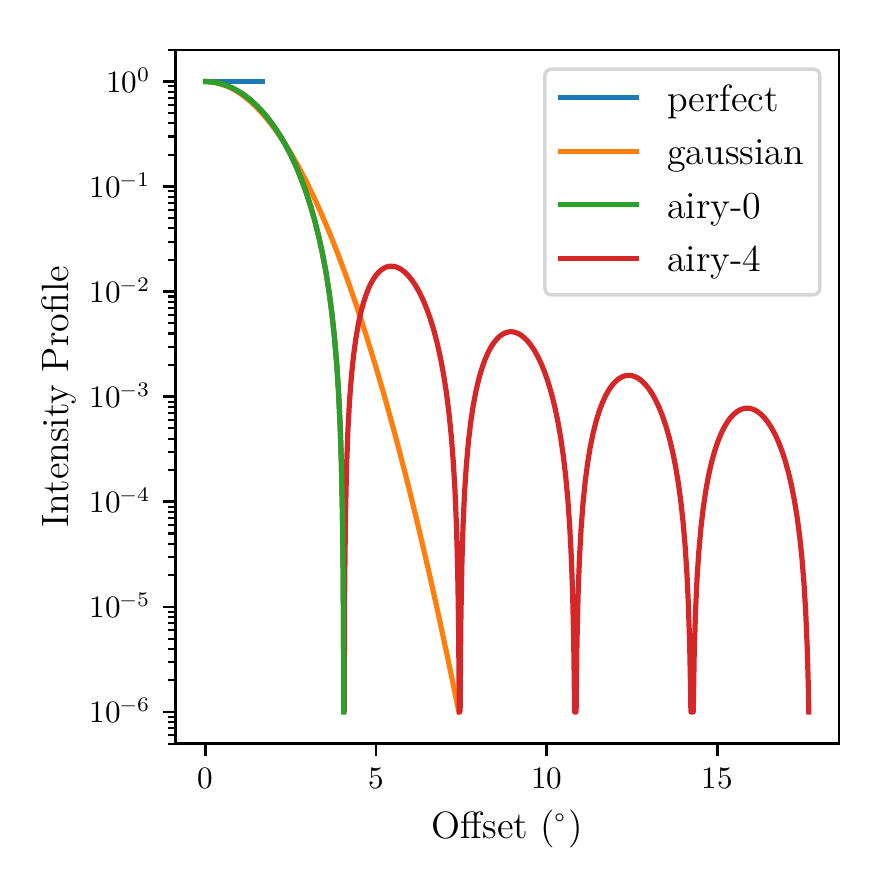}
 \caption{Plot showing the intensity profile of various beam patterns as function of the radial offset from the centre. The relative scaling on the vertical axis is linked to selected survey's beam size at \FWHM, as calculated from beamsizes seen in Table~\ref{tab:surveys}. Up to eight sidelobes can be included in \frbpoppy surveys, but the option to simulate a beam out to the \FWHM is also possible (as illustrated by the `perfect' beam pattern).}
 \label{fig:int_pro_theory}%
\end{figure}

In Fig.~\ref{fig:int_pro_theory}, examples of these beam patterns can be seen including several sidelobe options for an Airy disk. Note that in the latter cases a sidelobe of 0.5 can be chosen to cut the intensity profile at the full width at half maximum \FWHM. The choice of sidelobe sets the maximum radius at which an FRB can still be detected. The difference in sky area covered by an Airy disk without sidelobes and an Airy disk with 8 sidelobes is accounted for within \frbpoppy by recalculating the associated beam size. \\

\emph{Parkes}
Using the beam pattern described in \citet{parkesbeam} with an applied scaling between 0-1, an FRB can be randomly dropped in the calculated beam pattern, allowing for a more realistic intensity profile model when attempting to reproduce Parkes detections. This beam pattern uses the `MB21' setup combining 13 beams spanning 3x3 degrees on the sky, and is calculated at 1357~MHz, close to the central frequency adopted for Parkes in this survey.\\

\emph{Apertif}
In a similar fashion to the Parkes beam model, we can use the intensity profile developed for Apertif
(K.~Hess, \emph{priv.~comm.}; \citealt{2019NatAs...3..188A}).\\

\begin{figure}
 \centering
 \includegraphics[width=\linewidth]{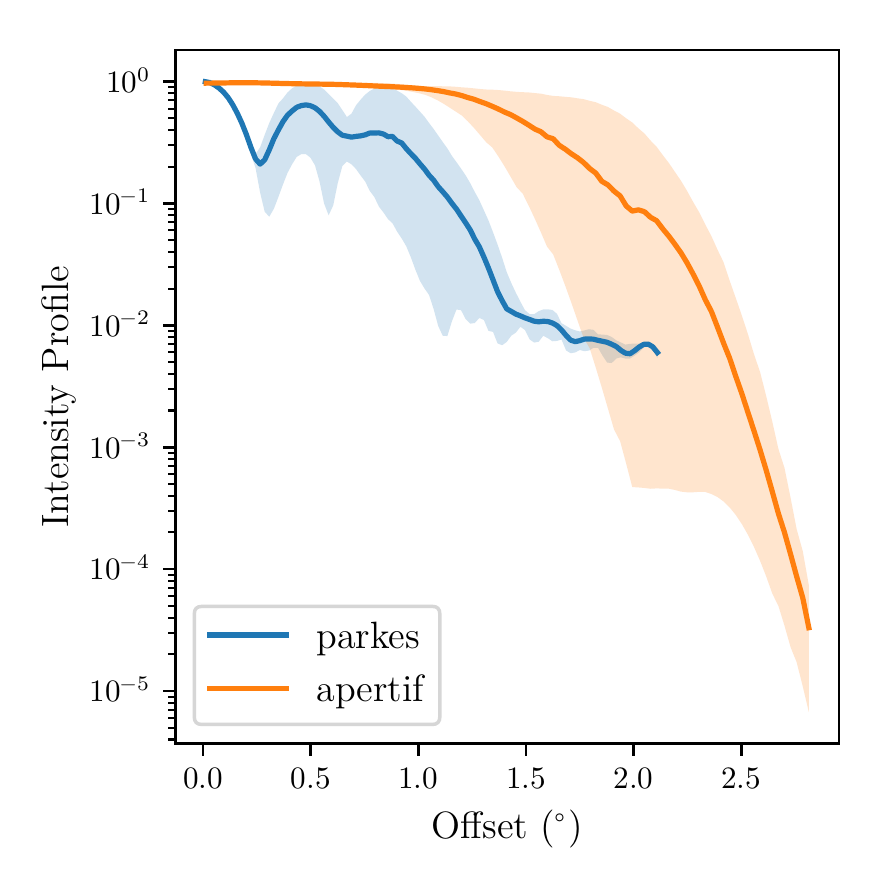}
 \caption{In the shaded regions possible beam intensities of respectively the Parkes Multibeam and Apertif Phased Array
  Feed (PAF) as a function of radial offset from the centre of the beam are shown (\citealt{parkesbeam}; K.~Hess, \emph{priv.~comm.}). Solid lines denoted the
  average intensity profile per survey.
  \label{fig:int_pro_surveys}%
 }
\end{figure}

In Fig~\ref{fig:int_pro_surveys}
we show the distribution of intensity profiles for both the Parkes and Apertif beam.
Shaded regions depict the range of intensities per radius, the darker lines indicate the average intensity profile.

\subsection{Rates}
We first determine the registered FRB detections by a survey's S/N limit (see Table~\ref{tab:surveys}). The rate at which FRBs are detected from a given redshift is, however, additionally affected by cosmological time dilation. To account for this effect, \frbpoppy dilutes the rate of detection by only recording a subset of events from redshift $z$, with this fraction being equal to to 1/$(1+z)$. This is done by drawing a random number $r \in [0,1]$ and testing for $r \leq (1+z)^{-1}$. Should an FRB satisfy this requirement, it is registered as detected, else as too late for detection. This mimics the finite observing window of a real survey.

While \frbpoppy uses all detected FRBs ($n_{\rm det}$) e.g. in simulating observed distributions, it would not be realistic to assume all generated FRBs happened to land within the beam of the telescope. In order to obtain a realistic detection rate of FRBs $r_{\rm det}$, $n_{\rm det}$ must be scaled by total survey area.
This can  be scaled from $n_{\rm det}$ with
\begin{equation}
 r_{\rm det} = \left( \frac{n_{\rm det}}{n_{\rm days}} \right) \left( \frac{A_{\rm beam}}{A_{\rm survey}} \right)
\end{equation}
with the detection rate $r_{\rm det}$, the number of detected FRBs $n_{\rm det}$, the number of surveying days $n_{\rm days}$, the FoV $A_{\rm beam}$, and the size of the survey area $A_{\rm survey}$. Here the number of surveying days $n_{\rm days}$ has been introduced to be able to discuss the detection rate of a single survey. Should the detection rates of multiple surveys be compared, this term could removed by normalising detection rates to that of a single survey. As
\begin{equation}
 A_{\rm survey} \simeq \frac{n_{\rm survey\ area}}{n_{\rm tot}} A_{\rm sky}
\end{equation}
in the limit of large $n$ and with $n_{\rm survey\ area}$ all FRBs within the survey area, detected or not. Using
\begin{equation}
 A_{\rm sky} = 4 \pi \left(\frac{360}{2\pi}\right)^2
\end{equation}
$r_{\rm det}$ can be calculated as
\begin{equation}
 r_{\rm det} = \left( \frac{n_{\rm det}}{n_{\rm days}} \right) \left(\frac{n_{\rm tot}}{n_{\rm survey\ area}} \right) \left( \frac{A_{\rm beam}}{A_{\rm sky}} \right)
\end{equation}
with the detection rate $r_{\rm det}$, the number of detected FRBs $n_{\rm det}$, the number of surveying days $n_{\rm days}$, the number of simulated FRBs $n_{\rm tot}$, the number of FRBs falling within the survey area $n_{\rm survey\ area}$, the FoV $A_{\rm beam}$, and the size of the survey area $A_{\rm survey}$.

Note however that this equation only holds for a population of one-off FRB events. While there now are two known repeating FRBs \citep{r1,r2}, the majority of the FRBs in the total population have only been seen once. Recent work by for instance \citep{ravi2019prevalence} seems to favour most observed FRBs to be originating from repeaters. Given the limited understanding of the repeating FRBs found so far, we have however chosen to model FRBs in \frbpoppy as single one-off events in this paper, and focus on repeaters in future work.

\subsection{Running \frbpoppy}
Using a set up as described in the sections above, \frbpoppy is able to construct a cosmic population with population parameters given in Table~\ref{tab:populations}. Subsequently a survey can be modelled using the survey parameters in Table~\ref{tab:surveys}. Convolving these two allows a survey population to be simulated. A minimum working example is given below, showing how \frbpoppy can be used:

\begin{minted}[fontsize=\scriptsize]{python}
# Import frbpoppy
from frbpoppy import CosmicPopulation, Survey, SurveyPopulation, plot

# Set up populations
cosmic_pop = CosmicPopulation(1e5)
survey = Survey('HTRU')
survey_pop = SurveyPopulation(cosmic_pop, survey)

# Check populations
print(survey_pop.rates())
plot(cosmic_pop, survey_pop)
\end{minted}

While this shows a basic setup, a large range of parameters can be given as arguments to these classes, providing the option for a user to tweak populations to their preference. The first run of \frbpoppy for a population of this size will typical take ${<}2$h on a 4 core computer, and will create databases for cosmological and dispersion measure distributions. Subsequent runs will be in the order of seconds. Increasing the population size to $10^8$ FRBs on a single core increases the run time to just over 3h, of which most time is spent on SQL queries to the generated databases.

\section{Forming a real FRB population}
\label{sec:real}
Real observations are needed to compare our simulations to reality. This section describes the process in which real data is gathered for use within \frbpoppy, from FRB parameters to detection rates.

\subsection{FRB parameters}
To verify simulated FRB distributions, \frbpoppy needs real FRB detection survey data.
To this end we use  FRBCAT, the online catalogue of FRBs\footnote{www.frbcat.org} \citep{frbcat}.
Some simple cleaning and conversion algorithms are applied to the database before use.
To obtain a single range of parameters per FRB, we filter the FRBCAT sample by selecting the measurement with the most
parameters. By default, repeat pulses are also filtered out to reduce the saturation of distributions by a single FRB
source. We subsequently attempt to match all FRBs with an associated survey using a user-predefined list.
\frbpoppy updates its database monthly if on line, and otherwise uses the most recent database. In this paper, all results have
been run using FRBCAT as available on 23 Sep 2019. Next to the entire real FRB population, \frbpoppy provides the option
to select FRBs from a single survey or telescope. An interactive plotting window can compare chosen populations.

\subsection{FRB detection rates}
Beyond the  parameters of individual FRBs, described above,
the \emph{rate} of detection is important to constrain the intrinsic FRB population.
Survey detection rates are not always published, often due to the difficulties in
determining the total observing time.

For the surveys that did publish rates, we convert the published rates into rates per survey expressed as the number of FRBs detected per day of observing time. In this paper we adopt

$R_{\rm htru}\sim
 0.08$~FRBs/day \citep{champion}, $R_{\rm askap-fly}\sim 0.12$~FRBs/day \citep{askap_nature} and $R_{\rm palfa}\sim
 0.04$~FRBs/day \citep{patel} and  $R_{\rm utmost}\sim 1/63$~FRBs/day \citep{utmost}.
 These rates encapsulate limits by their survey nature, whether in terms of observing frequency, fluence thresholds, sky coverage or any other selection effects. With \frbpoppy we expect to reproduce these rates, by virtue of replicating the underlying selection effects.
These rates are based on the highest estimated total time each survey was at full sensitivity; so actual detection rates could be lower.

\section{Comparing the simulated and observed FRB populations}
\label{sec:comp}
Ideally, simulated FRB populations can reproduce observed FRB populations. To this end, methods are needed by which populations can be compared. The following sections describe a number of these methods.

\subsection{FRB detection rates}
\label{sec:comp:rate}
Comparing simulated and real detection rates provides a first measure by which a simulation can be judged. With FRB detections expected to follow a Poissonian distribution, we take care to compare simulated detections to real ones within Poissonian error margins. With higher detection numbers providing stronger constraints on detection rate, surveys with more detections will necessarily show tighter constraints on acceptable simulated detection rates.

\subsection{FRB parameters}
\label{sec:comp:par}
We quantify the goodness of our model by producing an ensemble likelihood over the parameters we find most important: the distributions of dispersion measure and fluence.
For each model run, we take the product of the KS test values of these two parameters. This approach is one of the standards in pulsar population synthesis.
More parameters could be easily be included in this approach, allowing a user to focus on particular parts of the parameter space. Although the current work only explores certain individual survey populations, this defined goodness-of-fit allows us in principle to automatically explore the higher-dimensional parameter space to find the best representation of the true FRB population.
While in this work we use the dispersion measure and fluence to ascertain the goodness-of-fit, other parameters are also stored. Table~\ref{tab:output_params} shows a selection of the parameters available as part of a simulated survey population. We provide an interactive tool within \frbpoppy to compare all parameters between FRB survey populations, simulated or real.

\begin{table}
 \caption{Selection of parameters available within \frbpoppy for a surveyed FRB population. Note the parameter space is
   not fully independent, with several parameters dependent on each other.
  \label{tab:output_params}
 }
 \centering
 \begin{tabular}{l l}
  \hline\hline
  Parameters                                    & Units                   \\
  \hline
  Comoving distance                             & Gpc                     \\
  Redshift                                      & -                       \\
  Right ascension / declination                 & \degr                   \\
  Galactic longitude / latitude                 & \degr                   \\
  Bolometric luminosity                         & ergs/s                  \\
  Dispersion measure (total/host/IGM/Milky Way) & $\text{pc}/\text{cm}^3$ \\
  Signal to noise ratio                         & -                       \\
  Peak flux density                             & Jy                      \\
  Pulse width (effective / intrinsic)           & ms                      \\
  Fluence                                       & Jy ms                   \\
  Spectral index                                & -                       \\
  \hline
 \end{tabular}
\end{table}

\section{Results}
\label{sec:results}
\subsection{log N -- log S}
\begin{figure}
 \centering
 \includegraphics{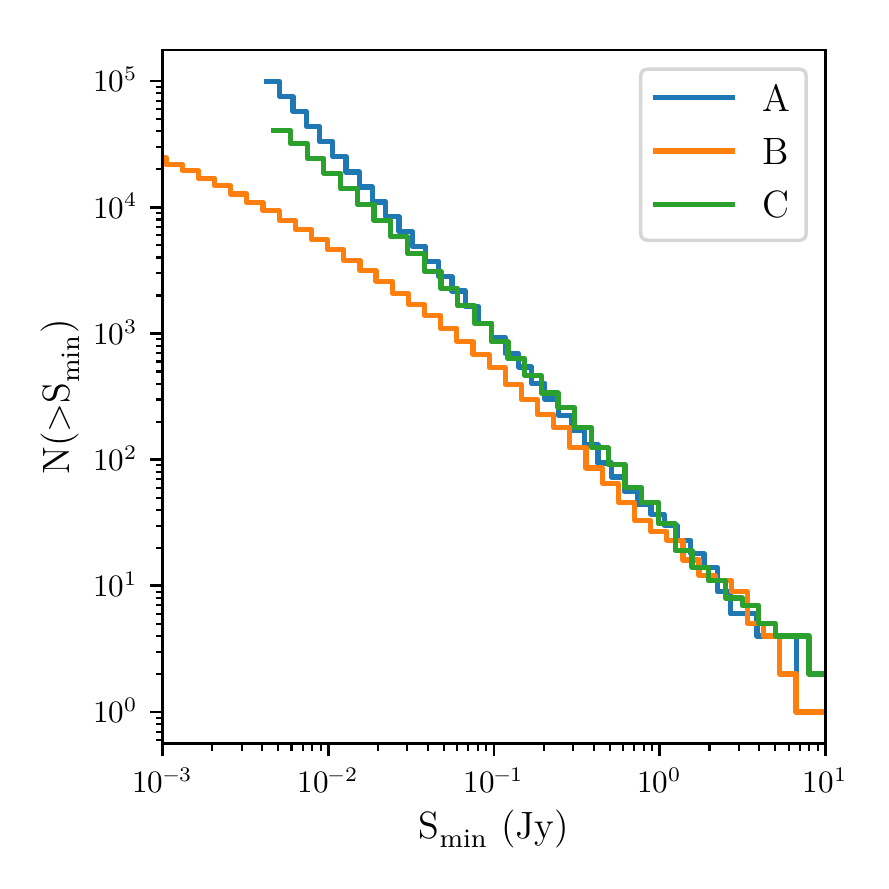}
 \caption{Cumulative source counts distribution of the number of detected FRBs greater than a limiting minimum detectable peak flux density, or $\log N$--$\log S$ plot. The resulting plots for three populations are shown here, with  ${\rm pop}_{\rm A}(z_{\rm max}, L_{\rm bol}, \gamma) = (0.01, 10^{38}, 0)$, ${\rm pop}_{\rm B}(z_{\rm max}, L_{\rm bol}, \gamma) = (2.5, 10^{42.5}, -1.4)$ and ${\rm pop}_{\rm C}(z_{\rm max}, L_{\rm bol}, \gamma) = (2.5, 10^{43}, 1)$. Despite all three populations probing very different parts of the universe, it is clear they exhibit very similar detection parameters at high fluxes. As such, this figure is used as an illustrative example of the danger in trying to interpret the underlying intrinsic FRB population from  a single $\log N$--$\log S$  plot. }
 \label{fig:logn_logs_abc}%
\end{figure}

The FRB source population has a sizeable number of parameters whose values are not well known
(see Table \ref{tab:populations}). Trying to infer properties of the cosmic population from a single histogram may be tempting, but we do not find it constraining.
An example of the risks can be seen in Fig.~\ref{fig:logn_logs_abc} in which a $\log N$--$\log S$ plot is
shown for three distinct and very different populations.
In this plot, population $A$ is the observed brightness distribution for a local population of standard candles with a flat spectral index. Population $B$ and $C$ go out to a larger redshift, with necessarily-higher luminosities and varying spectral indices such that

\begin{align}
 {\rm pop}_{\rm A}(z_{\rm max}, L_{\rm bol}, \gamma) & = (0.01, 10^{38}, 0) \nonumber \\
 {\rm pop}_{\rm B}(z_{\rm max}, L_{\rm bol}, \gamma) & = (2.5, 10^{42.5}, -1.4)       \\
 {\rm pop}_{\rm C}(z_{\rm max}, L_{\rm bol}, \gamma) & = (2.5, 10^{43}, 1) \nonumber
\end{align}

These simulated populations have been detected with a \survey{perfect} survey setup, allowing for instrumental effects to be decoupled from the observed source counts. \citet{amiri2017} emphasised the fact that for cosmological populations, the brightness distribution of FRBs is not expected to be described by a single power-law, though almost all brightness distributions should asymptote towards the Euclidean scaling at high flux densities.
Fig.~\ref{fig:logn_logs_abc} demonstrates the expected behaviour;  distinctions can be made between the three
populations at low flux densities. On the other end, in the limit of high flux densities, these populations have similar slopes
despite having very distinct intrinsic properties. While for instance plotting the spectral indices would discriminate between these populations, in the limit of high flux densities a $\log N$--$\log S$ plot by itself can not.
Fig.~\ref{fig:logn_logs_abc} serves both as a verification of \frbpoppy and
as a cautionary tale for trying to interpret the underlying intrinsic FRB population from just a single
distribution. This validates our use of careful population synthesis, and of using a multi-dimensional goodness-of-fit.

\subsection{Event rates}
While our models can be quite complex in general,
particular conditions exist that simplify them,
and allow for direct comparison to  analytical expectations.
This provides a way to test our code and assumptions.
As first metric for such a test, we take the  detection rates \frbpoppy surveys produce.
These can be tested against rather straight-forward analytical scaling relationships. \citet{connor2016b} show how the relative FRB detection rates of surveys $A$ and $B$ observing in a similar band can be expressed using the slope of the source count distribution:
\begin{equation}
 \frac{R_A}{R_B} = \frac{\Omega_A}{\Omega_B} \left( \frac{\text{\SEFD}_A}{\text{\SEFD}_B} \frac{\text{S/N}_{A}}{\text{S/N}_{B}}\right)^{\alpha_{\rm in}} \left( \frac{\Delta \nu_A}{\Delta \nu_B}\right)^{-\alpha_{\rm in}/2}
 \label{eq:scaling}
\end{equation}
each with a detection rate $R$, Field-of-View $\Omega$, system equivalent flux density \SEFD,  minimum signal to noise S/N, bandwidth $\Delta \nu$, and assuming an intrinsic slope of the source count distribution $\alpha_{\rm in}$ (see Sect.~\ref{sec:alpha}).

\begin{figure*}
 \centering
 \includegraphics{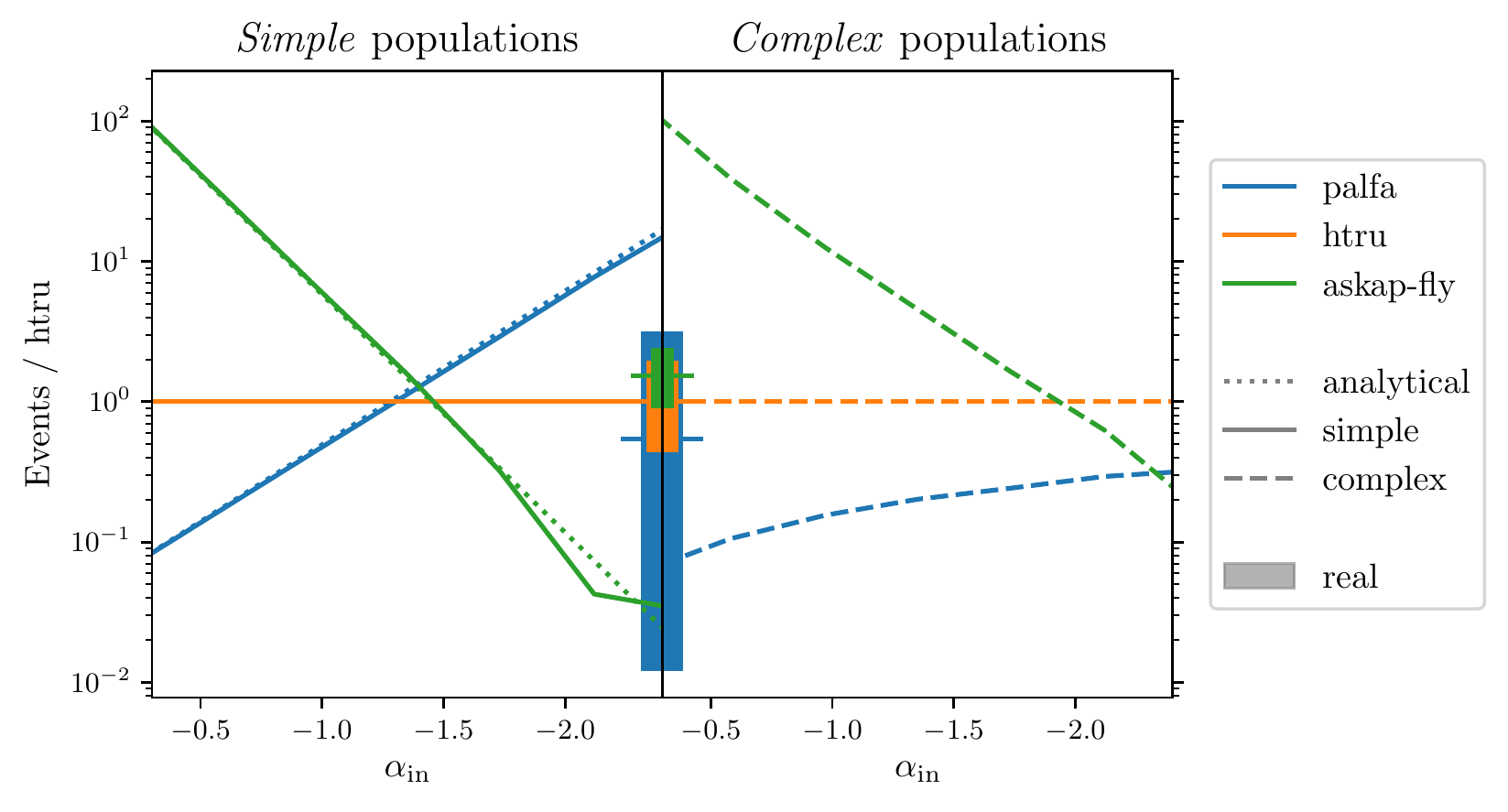}
 \caption{ The relative detection rates of three surveys as a function of the source counts slope, $\alpha_{in}$.
    Detection rates are normalized to the HTRU rate, using a Euclidean Universe with standard candles (\pop{Simple} population,
    \emph{left} panel), and a cosmological population with a broad
    luminosity function (\pop{Complex} population, \emph{right} panel). The dotted
    line is computed analytically; solid and dashed lines are the results of \frbpoppy. The real detection rate per survey is giving in the centre, with solid blocks denoting 1$\sigma$ Poissonian error bars (\pop{real} population, \emph{centre}). }
 \label{fig:event_rates}
\end{figure*}

These scaling relationships should hold for a local, non-cosmological population, as their source counts are expected to be given by a single power-law. If the brightness distribution is not well described by a single power-law, the relationship between sensitivity and detection rates becomes more complicated. For example, a sensitive telescope like
Arecibo would have an advantage over less sensitive telescopes if FRBs are described by population $C$ instead of $B$ in Fig.~\ref{fig:logn_logs_abc}. This is because the relative number of events falls off at low fluences in population $B$ as the slope of the source counts flattens. Therefore, we should find that surveys probing lower fluences would see fewer FRBs than the analytical relationship would predict for a population like $B$.
Additionally it can help to set FRB sources to be standard candles to ensure a similar volume is probed by both surveys. Finally, using a perfect beam pattern rather than an Airy disk prevents any beam pattern effects from playing a role in the relative FRB detection rates. Combining these premises into a \pop{Simple} intrinsic population (see Table~\ref{tab:populations}) and surveying this population with a range of surveys allows detection rates at various values of $\alpha_{\rm in}$ to be compared to the analytical expectations from Eq.~\ref{eq:scaling}. Such a comparison is made in Fig.~\ref{fig:event_rates} for \survey{palfa} and \survey{askap-fly} relative to those of \survey{htru} as a function of $\alpha_{\rm in}$. The expected analytical relationship is shown in dotted lines, with the results from \frbpoppy overplotted in solid lines.

The simulated results from the \pop{Simple} model  match the analytical expectations very well, showing \frbpoppy acts as expected within
understandable conditions. Furthermore, the change in detection rate over $\alpha_{\rm in}$ for \survey{palfa} agrees with prior expectations from
\citet{amiri2017}. The slight deviation from the trend around $\alpha_{\rm in}=-2.1$ for \survey{askap-fly} is solely due to insufficient detections, with larger populations eliminating this effect. Based on these results from these test cases we conclude that generating and surveying FRB populations \frbpoppy
works as expected. This paves the way for more complex behaviour to be tested, as done below.

One metric that is influenced by important and diverse elements such as the source number density,
the luminosities, and the telescope modelling, whether in sensitivity, beam pattern or other detection parameters, is the detection rate.
Thus comparing simulated detection rates to real ones is an important test of our population synthesis.
To this end, the
real detection rates of \survey{palfa}, \survey{htru} and \survey{askap-fly} have been plotted in the centre of
Fig.~\ref{fig:event_rates} using short horizontal lines.
The surrounding blocks denote the first order Poissonian error
bars for each survey.
These real detection rates can be used to constrain expected detection rates, and hence the underlying number density slope.
The left panel of Fig.~\ref{fig:event_rates} makes clear that even with simple analytic models, and a simple
and well defined source-count falloff such that $\alpha = \alpha_{\rm in}$, in $1.3<|\alpha|<1.5$ the observed FRB rates of the three main surveys are reproduced.
We take this, and the replication of the analytical expectations, as evidence that the fundamental simulation and detection numbers in \frbpoppy are correct and trustworthy.

We subsequently move to the more physically meaningful regime, leaving behind the oversimplification of the \pop{Simple} population, and shifting to a \pop{Complex} intrinsic FRB population.
The effects of adopting this population can be seen in right panel of Fig.~\ref{fig:event_rates}. The dashed lines denote the detection rates for a
variety of simulated surveys. In Table~\ref{tab:populations} we provide an overview of the initial input parameters for
this \pop{Complex} population. Having stepped away from a \pop{Simple} population, the interpretation of $\alpha_{\rm in}$ also changes. As described in Sect.~\ref{sec:alpha}, $\alpha_{\rm in}$ is only equal to the slope of $\log N$--$\log S$ $\alpha$ for a Euclidean universe, in all other cases $\alpha_{\rm in}$ become the value to which $\alpha$ asymptotes in the limit of high fluences.

We choose to model the \pop{Complex} population by including dispersion measure contributions, a range of luminosities rather than a standard candle, and also a negative spectral index similar to the Galactic pulsar population. Additionally, we adapted surveys to use Airy disk beam
patterns with a single sidelobe.

\begin{figure}
 \centering
 \includegraphics[width=\columnwidth]{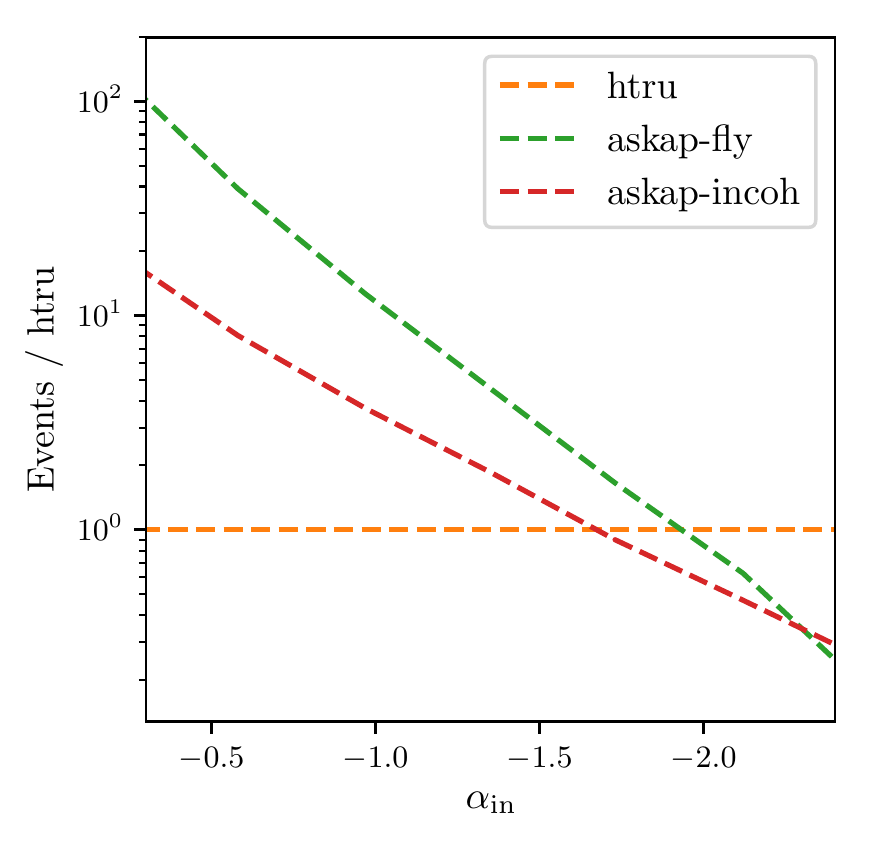}
 \caption{Simulated relative detection rates for ASKAP in Fly's Eye mode (\survey{askap-fly}) and in an incoherent mode (\survey{askap-incoh}) scaled to \survey{htru} and plotted against the source counts slope, $\alpha_{in}$. A \pop{Complex} model of the intrinsic FRB population has been used, with both surveys modelled with an Airy disk with a single sidelobe.}
 \label{fig:askap_rates}%
\end{figure}

Looking at Fig.~\ref{fig:event_rates} we find the relative \survey{askap-fly} / \survey{htru} detection rates increase while the \survey{palfa} detection rate drops significantly at high values of $\alpha_{\rm in}$ and loosens its constraints. As a result, the expected range for $\alpha_{\rm in}$ is pushed towards $1.5<|\alpha_{\rm in}|<2.0$. Such a range of values for $\alpha_{\rm in}$ correspond a value for $\beta<1$ (see Eq.~\ref{eq:vol_co_pl}). Therefore we expect the comoving FRB source density to drop off towards higher redshift, indicating an evolution in the number of FRB sources. It implies FRB sources in the early universe were less common than in the later stages of the universe which could help in tying down the FRB progenitors to an astrophysical source class.
Extending such simulations to \survey{askap-incoh} can be used to predict the expected change in ASKAP detection rates. This is shown in Fig.~\ref{fig:askap_rates}, using a similar setup to the right panel of Fig.~\ref{fig:event_rates}. In this case the choice is made to limit the surveys to \survey{htru}, \survey{askap-fly} and \survey{askap-incoh}, of which more details can be seen in Table~\ref{tab:surveys}. Comparisons to real rates can be made using Fig.~\ref{fig:event_rates}, which are additionally applicable to Fig.~\ref{fig:askap_rates}.

\subsection{Distributions}
A crucial first step for any simulation is its ability to replicate the observed results;
the second step is adjust the input model to maximise the quality of this replication and thus understand
the input astrophysics.
Our replicated parameter space includes many variables, as described in Sect.~\ref{sec:comp:par}.

\begin{figure}
 \centering
 \includegraphics[width=\columnwidth]{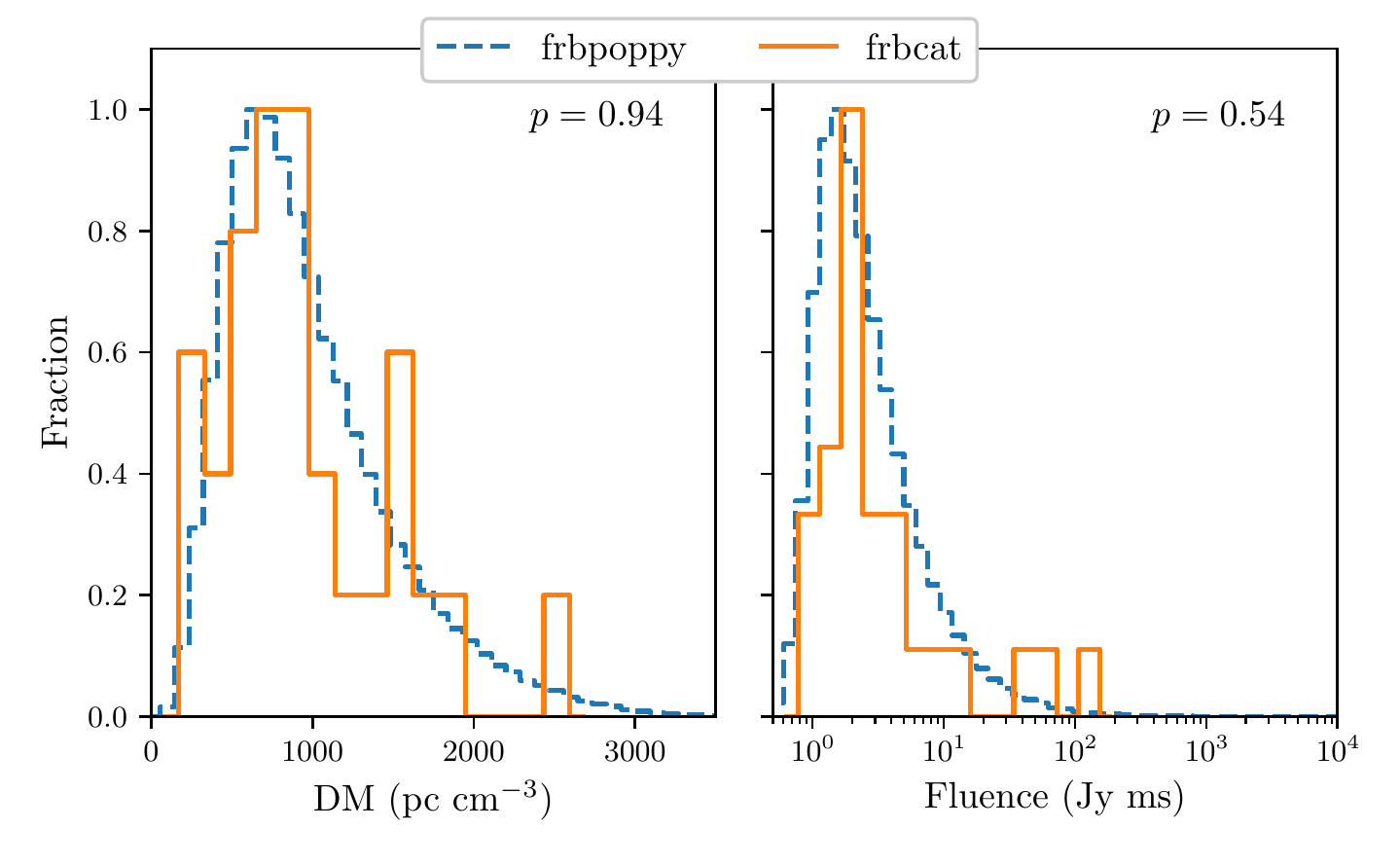}
 \caption{Comparison of simulated \frbpoppy and real frbcat distributions for FRB detections at Parkes. (\emph{left}) Dispersion measure distributions (\emph{right}) Fluence distributions for the same populations as the left-hand panel.
 \frbpoppy simulations have been run on a \pop{Complex} intrinsic FRB population with the \survey{parkes} survey modelled using the beam pattern as shown in Fig.~\ref{fig:int_pro_surveys}. The $p$-value of a simple KS-test between both distributions can be seen in the upper right corner of both panels. The product of these values showing the total goodness-of-fit is $p=0.51$. Note the chosen input parameters do not reflect the optimum values for the best fits between \frbpoppy and frbcat distributions, but merely an initial guess at some of the underlying parameters.}
 \label{fig:frbpoppy_parkes}%
\end{figure}

\begin{figure}
 \centering
 \includegraphics[width=\columnwidth]{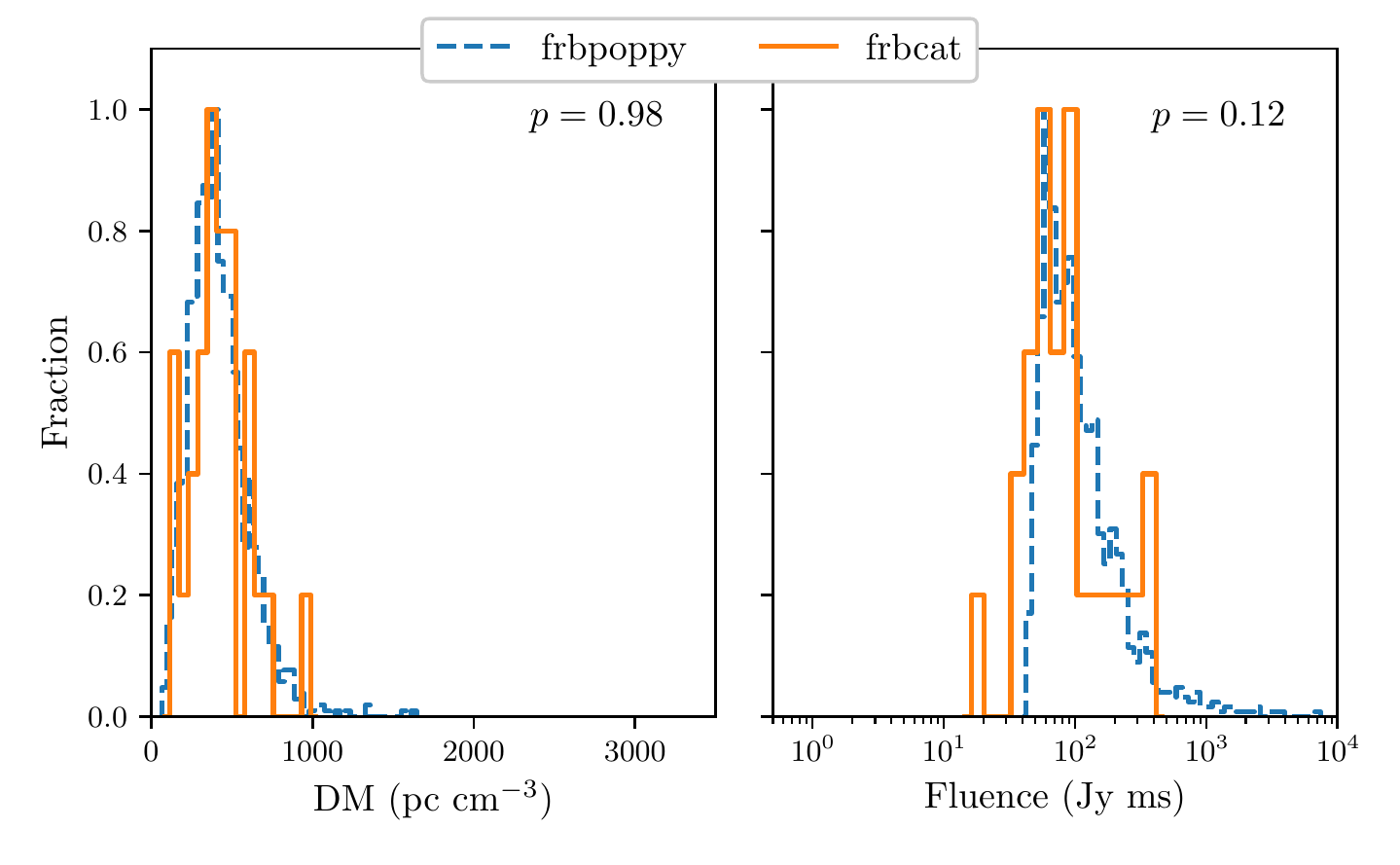}
 \caption{Following Fig.~\ref{fig:frbpoppy_parkes}, \frbpoppy and frbcat dispersion measure and fluence ASKAP distributions. The \frbpoppy populations use a \pop{Complex} intrinsic FRB population and are surveyed with \survey{askap-fly} and including a single sidelobe. The total goodness-of-fit is $p=0.12$.}
 \label{fig:frbpoppy_askap}%
\end{figure}

Here we show a comparison of just two parameters - fluence and dispersion measure -- which provide a rough measure of brightness and distance, respectively.
We compare simulated and observed fluence and dispersion measure distributions from Parkes and ASKAP. The resulting plot for Parkes can be seen in Fig.~\ref{fig:frbpoppy_parkes} and for ASKAP in Fig.~\ref{fig:frbpoppy_askap}. In order to obtain these results we surveyed a \pop{Complex} intrinsic FRB population with a \survey{parkes} survey using the Parkes beam pattern, and with an \survey{askap-fly} model using an Airy disk with a single sidelobe. More details on the intrinsic population can be found in Table~\ref{tab:populations}, with information on the survey parameters in Table~\ref{tab:surveys}, and an idea of the Parkes beam pattern in Fig.~\ref{fig:int_pro_surveys}.

Fig.~\ref{fig:frbpoppy_parkes} and Fig.~\ref{fig:frbpoppy_askap} show the broad trends of the resulting \frbpoppy
distributions to be quite similar to those from FRBCAT, with a KS-test values of $p=0.51$ for the Parkes results, and $p=0.12$ for ASKAP. Not only does this support \frbpoppy's capability to reproduce observed data, but it shows the \pop{Complex} population parameters to be a favourable place from which to explore the intrinsic population parameter space. In comparison, a \pop{Simple} population for instance was unable to reproduce observed fluence and dispersion measure distribution. Comparing the inputs to the two populations as given in Table~\ref{tab:populations} shows a number of key differences. A number of parameters proved crucial for replicating real detections. Both the cosmological nature of the \pop{Complex} population, in obtaining a good match to observed data, as the lognormal nature of the pulse width distribution proved to be important factors. This shows that the intrinsic FRB population to be more complex and varied than albeit tempting simple approximations of the intrinsic population.
Note the sampling difference in between \frbpoppy and FRBCAT - the latter comprises of tens of FRBs, with \frbpoppy showing hundreds.  With additional real FRB observations the constraints on the intrinsic FRB population could be tighter.

\subsection{Beam patterns}
\label{sec:results:beams}
\begin{figure}
 \centering
 \includegraphics{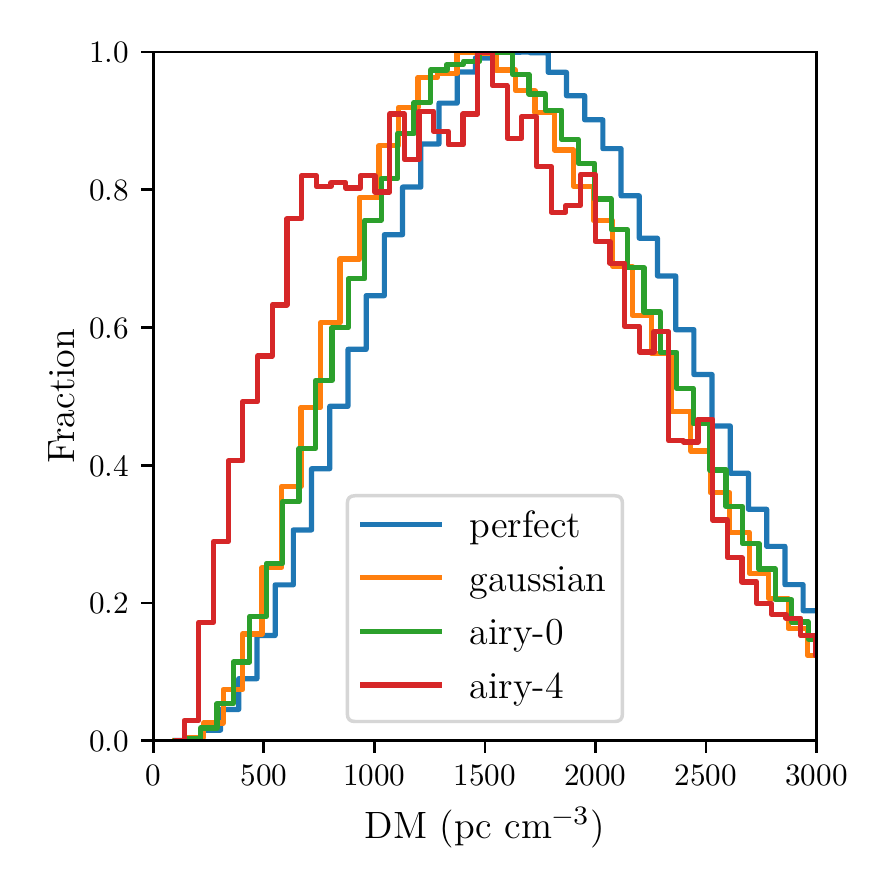}
 \caption{The relative fraction of FRB detections over dispersion measure for a variety of beam patterns. In this case airy-0 and airy-4 respectively denote an Airy disk with 0 sidelobes and one with 4. FRBs were simulated with the \textit{Standard Candles} class parameters.}
 \label{fig:dm_beams}%
\end{figure}

In general, the telescopes simulated in this work are most sensitive at boresight, and well understood there.
But away from this beam centre the sensitivity of an observation can be quickly reduced
as seen in Fig.~\ref{fig:int_pro_theory}, and the exact shape of the fall-off becomes important. Indeed, the beam pattern of telescopes such as Parkes isn't well known at large angular distances from the boresight.
One could imagine that adopting for instance an Airy disk with a large number of sidelobes might skew any resulting
distributions towards brighter FRBs with dim FRBs less likely to be detected. In Fig.~\ref{fig:dm_beams}, an example is
shown of the change in observed DM distributions based on the choice of beam pattern.
Where, with a perfect beam pattern, the simulated observed distribution is found peaking towards higher DM values, an
Airy or Gaussian profile shifts the peak leftwards, to lower DMs. More noticeable is the left shoulder of the Airy disk with 4 sidelobes, seeming to suggest a far steeper build up of FRB sources at low DMs despite the 'perfect' beam showing elsewise. If beam pattern effects are not properly taken into account, they will easily lead to erroneous conclusions about the intrinsic number density of FRBs.
Additionally such behaviour could complicate comparisons between surveys, each having their own, unique beam pattern effect convolved within their detections.
In Fig.~\ref{fig:dm_beams} the input parameters were chosen to best illustrate these effects, using a \pop{Standard
 Candles} population (see Table~\ref{tab:populations}) being observed with a perfect telescope setup (see
Table~\ref{tab:surveys}).
This survey was adapted to feature a smaller FoV of 10~deg$^2$ and detections made if
the peak flux density $S_{\text{peak}} > 10^{-10}$ Jy. Shifting the detection threshold causes the effects seen in Fig.~\ref{fig:dm_beams} to become less noticeable, though nonetheless still present.

\section{Discussion}
\label{sec:discussion}

\subsection{Caveats}

\subsubsection{Repeaters}
As described in Sect.~\ref{sec:generate},
this first version of \frbpoppy models FRBs as one-off events, even though repeating FRBs have been detected.
So why was this choice made?
Firstly, it would be difficult to do population statistics including repeaters considering at this stage only a handful
of repeaters are known. Secondly, should both  a repeater and a true one-off population underlie the observed FRBs, then
our results should still hold for the one-off population. This assumes that there would be a way to distinguish between both populations, as otherwise  contamination between the two populations would prohibit separate modelling.
The potentially long repetition timescale of repeater sources may indeed allow modelling FRBs as one-off sources.
Then one just needs to take care to take into account the number of pulses emitted over an FRB lifetime
when converting the FRB number density (Sect.~\ref{sec:generate:density}) to a birth rate.
Nonetheless, repeaters are being included in the future version of \frbpoppy, and subject to further synthesis
research. This will allow for characterising the fraction of repeating to non-repeating FRB sources, and potentially their progenitor populations.

\subsubsection{Beam patterns}
As we have shown in Sect.~\ref{sec:results:beams},  determining the beam pattern is essential in understanding results
of any survey.
Actual beam patterns are rarely ideal, as is strikingly clear from the differences between
Figs.~\ref{fig:int_pro_theory} and \ref{fig:int_pro_surveys}.

Furthermore, cylindrical telescopes such as UTMOST or CHIME have complex, elongated beam patterns
\citep[cf.][]{2017PASA...34...45B}. Results in Fig.~\ref{fig:dm_beams} demonstrate the importance of knowing the survey beam pattern, for the effects on resulting detections can be important. To ensure FRB detections from various surveys can be compared against each other, it is important surveys release not just survey parameters, but also a map of their beam pattern. Doing so will significantly improve the constraints surveys can place on the intrinsic FRB population.

While for \emph{pulsar} population studies the beam pattern is in principle the same, the effects of the sidelobes are
generally  much more important in FRB studies.
As sidelobes generally rotate on the sky, their effects wash out
during the relatively long integrations
used in the periodicity searching for pulsars.
Then only the central, axisymmetric parts of the beam shape add to a detection.
For FRB and other single-pulse searches the instantaneous beam pattern, including any strong sidelobes, is more important.

\subsubsection{Fluence limits}
One of the strengths of population synthesis is in tracking down detection biases. One such bias, as described in
\citep{keane2015completeness}, is within the fluence space. Two FRBs could have the same fluence,
and yet only one might be detected if e.g.~ their pulse widths differ. Thus sampling the
fluences of an FRB population will show an incomplete picture. A common method to ensure some form of completeness is by
shifting to the $S_{\rm peak}$--$w_{\rm eff}$ space and use a fluence completeness-limit. By rewriting Eq.~\ref{eq:S/N},
one can decide to use FRBs only if above a particular constant
fluence and below a maximum $w_{\rm eff}$ \citep[see e.g. Fig.~2 of][]{keane2015completeness}. While this indeed can prevent fluence incompleteness, it thereby overlooks other FRBs.
We replicate this incompleteness in \frbpoppy by using a simple S/N detection threshold, as surveys often do. By using a S/N threshold rather than a fluence completeness limit the \frbpoppy survey detections show the bias in the fluence space, necessarily being an incomplete sampling of the parameter space. Knowledge of the underlying pulse width distribution would help map out the extent of the selection effects mentioned in this paragraph, but may not be achievable in the near future.

\subsubsection{Software selection effects}
In generating our simulated observed surveys, we take care to model a number
of boundaries to the FRB search parameter space.
For the telescope hardware system these are usually well described; for the software, this is not always the case.
Thus we do model in \frbpoppy the minimum sampling time, but not the maximum searched pulse widths, and we model the intrachannel DM smearing but not the search DM-step smearing. Some care with simulation inputs is therefore advised, to ensure simulated detections remain well within the bounds of any software selection effects.
In general, a number of  search-software selection effects exist that are beyond the scope of the current work.
Several research teams are pursuing a more thorough investigation of this FRB search-software completeness and bias in separate line of
research \citep{mh18,frb-olym}.

\subsection{Comparing population synthesis for FRBs with that for pulsars}
The current research, and \frbpoppy, follow from population synthesis work in the pulsar community through
e.g. the open-source  \texttt{psrpoppy} code.
While there are a number of similarities, these are offset by some intrinsic differences.
In both cases, large numbers of sources need to be generated.
In the pulsar population synthesis of \citet{ls10}, for example,
the generated population sizes in a run from a single parameter set is generally $10^{7}$ pulsars.
For about $5 \times 10^{5}$ of these, the full orbit through the galaxy was simulated, a task \frbpoppy does not need to perform.
Searching through multiple parameters generally runs on clusters (or very large servers).
An FRB population quickly starts running into intrinsic populations of $10^{8}$ FRBs.

One main difference is, however, that already when pulsar population synthesis research began, neutron stars were known to be the source class.
Furthermore, a significant number had been localised, and their distances determined, and thus, their intrinsic brightnesses were well understood.
In contrast, with FRBs we find a lack of understanding on the intrinsic emission properties.
This makes the parameter space over which FRBs have to be modelled significantly larger than that for pulsars.

A further key difference between FRB and pulsar population synthesis are their respective one-off and periodic burst properties.
In an all-sky survey, a pulsar that is always on will be detected most brightly in the pointing where  the main beam points closest to it.
For FRBs this is not so. Indeed, they are most likely not emitting in that optimally-directed pointing.
Most FRBs will burst while covered by the larger sidelobes.
And emitting FRB is thus more dependant on its  placement within a beam pattern, with only a single chance to
detect one-off sources. This also severely impacts the detection rates in comparison to pulsars, with FRBs going off
outside of a beam pattern gone for ever.

\subsection{Comparing \frbpoppy results with other FRB simulations}
We have investigated how to compare results from \frbpoppy with those from the population synthesis studies listed in Sect.~1.
Direct comparisons are hampered by the different scope of the simulations, and the rapidly changing datasets.
\citet{caleb2016}, for example,  focused on 9 HTRU events, with little other data being available at the time. For the generation of their FRB populations, a similar path to \frbpoppy was taken, testing several cosmological models, adopting a linear DM-$z$ relationship and a range of telescope selection effects.
A number of fundamental differences in approach exist too, however,
such as the treatment of the spectral index, and in adopting scattering relationships.
Nonetheless, similar results were obtained in terms of fluence and dispersion measure distributions, with \frbpoppy showing a slightly better fit to Parkes detections,
($p_{\rm \frbpoppy}=0.51$ versus $p_{\rm Caleb\ et\ al.}=0.03$).
If the \citet{caleb2016} research were extend to the current detections (beyond the scope of the current paper),
it could be more directly compared to  \frbpoppy.
Facilitating such comparisons is one of the drivers for making \frbpoppy open source.

\subsection{Event rates}
FRB event rates can be difficult to interpret \citep{connor2016b}. The rate at which an individual survey detects bursts is the simplest to calculate. It is more difficult to convert that number into an all-sky rate, as one must then know the telescope's beam pattern and the FRB brightness distribution rather well \citep{jp1}. Harder still is producing a volumetric event rate, because that requires information about the spatial distribution of FRBs, and without redshifts for large numbers of sources, the volume of space occupied by FRBs is degenerate with their repetition statistics and luminosity function. Of course, localisations such as those by ASKAP \citep[e.g.][]{bannister2019} help to determine a volumetric rate, providing a redshift and thereby a handle on the luminosity function etc.

While it is the most difficult to constrain, the volumetric rate is the most informative quantity because it contains
information about the progenitor population. Given the difficulty of inverting a detection rate into a rate on the sky
and then a volumetric rate, running a large MCMC simulation with \frbpoppy is the best approach to constraining
frequency at which cosmological FRBs are produced.
\frbpoppy handles beam effects and instrumental biases and given enough resources can answer the question `Which volumetric event rates are consistent with current data?'. Therefore, it is promising that we have already shown the consistency between both absolute detection rates in \frbpoppy and the relative event rates between surveys, e.g. Fig~\ref{fig:event_rates}. Additionally, current detection rates constraining $|\alpha_{\rm in}|>1.5$ for a \pop{Complex} population, point towards a possible evolution of FRB sources with more occuring per unit comoving volume in the nearby universe than in the distant universe.

\subsection{Observed distributions}
In Fig.~\ref{fig:frbpoppy_parkes} and \ref{fig:frbpoppy_askap}, we show our simulated parameter distributions to be in good agreement with the observed ones, for the \pop{Complex} population. The properties of this model therefore warrant further examination. Starting with the number density, Table~\ref{tab:populations} shows the population following a comoving volume density, rather than Star Formation Rate (SFR). This choice was made so that the intrinsic number density could vary with $\alpha_{\rm in}$, and does not de facto rule out other number densities from being able to fit the data.
The fits as shown in Fig.~\ref{fig:event_rates} indicate some form of evolution in the FRB progenitor population.
We cannot, however, rule out a Euclidean distribution with the current data.
We will further explore the evolution of FRB progenitors in future work. This might allow current detections to tie FRBs to a progenitor population.

Additionally, the simulated population extends out to a redshift of 2.5, which leads to a choice of intrinsic bolometric luminosity of $10^{39}-10^{45}$~ergs/s.
Varying both the maximum redshift and the luminosities can result in a similar population (see Fig.~\ref{fig:logn_logs_abc}), so setting one of the two parameters helps set the other when aiming for a representative outcome. In the case of the distributions discussed here, the simulated luminosities agree with \citet{yang2017}, who advocate for a luminosity around $L \sim 10^{43}\ {\rm erg\ s^{-1}}$ with a narrow spread. The chosen luminosity range subsequently inform the choice of intrinsic pulse widths, which were drawn from a log-normal distribution. In future \frbpoppy runs, information from repeater sources could help inform the choice of pulse width distributions. Additional constraints could be placed using the pulse width distribution of detected FRBs, provided the strength of the pulse broadening effects in the host galaxy and intergalactic medium are well understood.

The strength of pulse broadening effects ties into the choice of intergalactic dispersion measure. While initial research suggested first-order approximation of ${\rm DM}_{\rm IGM} \sim 1200z$ with redshift $z$ \citep{inoue2004}, more recent treatments tend towards a smaller scaling factor between 800-1000~pc/cm$^{-3}$ \citep{zhang_dm,keane2018,pol2019}. We chose for a simple relation of 1000~pc/cm$^{-3}$ and a ${\rm DM}_{\rm IGM}$ drawn from a Gaussian centred around 100~pc/cm$^{-3}$, until information from both repeater sources, and improved localisations further constrain these values. We acknowledge that a more accurate relation could be obtained using non-linear relationships \citep[see][]{batten2019}, and may indeed be important at the redshift of HeII ionisation. Implementing such relationships directly in \frbpoppy would however significantly increase the computation time, and some form of pre-optimisation would have to be done.

Difficulties in measuring a spectral index $\gamma$ make setting this value challenging. Currently set to follow the spectral index seen in pulsars, with a value of $-1.4$ \citep{bates2013}, a diverse range of predictions are present in the literature. Where for instance \citet{macquart2019spectral} argue for a steep negative spectral index, \citet{farah2019five} suggests a possible spectral turnover in line with \citet{ravi2019explaining}. Further muddling the idea of a spectral index are repeater observations presenting indications of FRBs to be emitted in emission envelops \citep[e.g.][]{gourdji2019sample, hessels2019frb121102}. Testing a variety of relationships between the FRB source energy and frequency would help in this regard, and could be taken into consideration in future work. In any case, additional FRB spectral index (or shape) measurements would help inform the choice of $\gamma$ within \frbpoppy. As seen in the limit of low fluences in Fig.~\ref{fig:logn_logs_abc}, the spectral index affects the brightness distributions, and can help distinguish between intrinsic populations.

There are two obvious avenues to explore in future work on the distributions generated by \frbpoppy: simulating more variations on the intrinsic populations, and expanding the number of parameters which are fit in the code. Both paths will improve constraints on the intrinsic FRB population. The increase in FRB detections from new surveys will also make for better comparisons, by constraining the physical parameter space occupied by the real population. It is clear that while our current inputs can explain the observed FRB population, \frbpoppy provides fertile ground for further constraining the intrinsic FRB population.

\subsection{Opportunities, uses, and future work}
The open-source nature of \frbpoppy is meant to encourage survey teams to update
their survey parameters and add descriptions of new search efforts. The main
goal, however, is to allow an open platform for FRB population synthesis, such
that research teams can analyse the impact of new discoveries.
These can range from new algorithms for generating populations, to new diagnostic plots for investigating FRB properties.

With the basic \frbpoppy functionality here demonstrated, our next goals are to
simulate the influence of a number of physical unknowns. We thereby aim to
investigate their effects on our simulated population, and from inverting the
real population, determine how important these physical unknowns are.

Immediately examples of these are whether the FRB birth rate follows the SFR or
is flat; what the fraction of repeating FRBs is; and how many FRBs are broad-band emitters.
All these will the strongly guided
by the continuing results from existing and new surveys.

\section{Conclusions}
\label{sec:conclusion}
We have developed \frbpoppy, an open-source Python package capable of conducting Fast Radio Burst population synthesis. Using this software we can replicate observed FRB detection rates and FRB distributions. \frbpoppy does this in three steps:
\begin{enumerate}
    \item \frbpoppy starts by simulating a cosmic population of one-off FRBs, for which a user can choose from a large range of options including models for source number density, cosmology, DM host/IGM/Milky Way, luminosity functions, emission bands, pulse widths, spectral indices as well as choices for the maximum redshift and size of the FRB population. These are merely a selection of the frontend options, with more options available within \frbpoppy.
    \item \frbpoppy then generates a survey, by adopting a beam pattern and by using survey parameters such as the telescope gain, sampling time, receiver temperature, central frequency, bandwidth, channel bandwidth, number of polarisations, FoV, S/N limit and any survey region limits.
    \item In the final step \frbpoppy convolves the generated intrinsic population with the generated survey to simulate an observed FRB population.
\end{enumerate}

Testing \frbpoppy shows the FRB detection rates of ASKAP, Parkes and Arecibo can be reproduced, as well as the observed fluence and dispersion measure distributions of ASKAP and Parkes. These observed results are replicated best by our `Complex model' (multiple DM contributions, range of luminosities, negative spectral index).
Overall this enables predictions to be made about the detection rates of future surveys, and about the intrinsic FRB population. We demonstrated the importance of understanding a survey's beam pattern by comparing the effects of various beam patterns. Future work will focus on auto-iteration over input parameters, on FRB repetition, and on further constraining the intrinsic FRB population.

\begin{acknowledgements}
 We thank Chris Flynn for carefully reading this manuscript and for providing valuable comment and suggestions. We also thank the participants of the 2017-2019 FRB meetings for the useful discussions. Thanks additionally go to Sarah
 Burke-Spolaor for initial survey parameter data. DWG acknowledges travel support from the Research Corporation for
 Scientific Advancement to attend the Aspen FRB meeting in 2017. The research leading to these results has received
 funding from the European Research Council under the European Union's Seventh Framework Programme (FP/2007-2013) / ERC
 Grant Agreement n. 617199 (`ALERT'); from Vici research programme `ARGO' with project number
 639.043.815, financed by the Netherlands Organisation for Scientific Research (NWO) ; and from
 the Netherlands Research School for Astronomy (NOVA4-ARTS).
 EP further acknowledges funding from an NWO Veni Fellowship.
 This research has made use of \texttt{numpy} \citep{numpy}, \texttt{scipy} \citep{scipy}, \texttt{astropy} \citep{astropy}, \texttt{matplotlib} \citep{matplotlib}, \texttt{bokeh} \citep{bokeh} and NASA’s Astrophysics Data System.
\end{acknowledgements}

\bibliographystyle{yahapj}
\bibliography{frbpoppy}

\begin{thebibliography}{}
\providecommand\natexlab[1]{#1}
\providecommand\JournalTitle[1]{#1}

\bibitem[{{Adams} \& {van Leeuwen}(2019)}]{2019NatAs...3..188A}
{Adams}, E.~A.~K., \& {van Leeuwen}, J. 2019,
  \href{http://dx.doi.org/10.1038/s41550-019-0692-4}{\JournalTitle{Nature
  Astronomy}, 3, 188}

\bibitem[{{Amiri} {et~al.}(2017){Amiri}, {Bandura}, {Berger}, {Bond}, {Cliche},
  {Connor}, {Deng}, {Denman}, {Dobbs}, {Domagalski}, {Fandino}, {Gilbert},
  {Good}, {Halpern}, {Hanna}, {Hincks}, {Hinshaw}, {H{\"o}fer}, {Hsyu},
  {Klages}, {Landecker}, {Masui}, {Mena-Parra}, {Newburgh}, {Oppermann}, {Pen},
  {Peterson}, {Pinsonneault-Marotte}, {Renard}, {Shaw}, {Siegel}, {Sigurdson},
  {Smith}, {Storer}, {Tretyakov}, {Vanderlinde}, {Wiebe}, \& {Scientific
  Collaboration20}}]{amiri2017}
{Amiri}, M., {Bandura}, K., {Berger}, P., {et~al.} 2017,
  \href{http://dx.doi.org/10.3847/1538-4357/aa713f}{\JournalTitle{\apj}, 844,
  161}

\bibitem[{{Astropy Collaboration} {et~al.}(2018){Astropy Collaboration},
  {Price-Whelan}, {Sip{\H o}cz}, {G{\"u}nther}, {Lim}, {Crawford}, {Conseil},
  {Shupe}, {Craig}, {Dencheva}, {Ginsburg}, {VanderPlas}, {Bradley},
  {P{\'e}rez-Su{\'a}rez}, {de Val-Borro}, {Aldcroft}, {Cruz}, {Robitaille},
  {Tollerud}, {Ardelean}, {Babej}, {Bach}, {Bachetti}, {Bakanov}, {Bamford},
  {Barentsen}, {Barmby}, {Baumbach}, {Berry}, {Biscani}, {Boquien}, {Bostroem},
  {Bouma}, {Brammer}, {Bray}, {Breytenbach}, {Buddelmeijer}, {Burke},
  {Calderone}, {Cano Rodr{\'{\i}}guez}, {Cara}, {Cardoso}, {Cheedella},
  {Copin}, {Corrales}, {Crichton}, {D'Avella}, {Deil}, {Depagne}, {Dietrich},
  {Donath}, {Droettboom}, {Earl}, {Erben}, {Fabbro}, {Ferreira}, {Finethy},
  {Fox}, {Garrison}, {Gibbons}, {Goldstein}, {Gommers}, {Greco}, {Greenfield},
  {Groener}, {Grollier}, {Hagen}, {Hirst}, {Homeier}, {Horton}, {Hosseinzadeh},
  {Hu}, {Hunkeler}, {Ivezi{\'c}}, {Jain}, {Jenness}, {Kanarek}, {Kendrew},
  {Kern}, {Kerzendorf}, {Khvalko}, {King}, {Kirkby}, {Kulkarni}, {Kumar},
  {Lee}, {Lenz}, {Littlefair}, {Ma}, {Macleod}, {Mastropietro}, {McCully},
  {Montagnac}, {Morris}, {Mueller}, {Mumford}, {Muna}, {Murphy}, {Nelson},
  {Nguyen}, {Ninan}, {N{\"o}the}, {Ogaz}, {Oh}, {Parejko}, {Parley}, {Pascual},
  {Patil}, {Patil}, {Plunkett}, {Prochaska}, {Rastogi}, {Reddy Janga},
  {Sabater}, {Sakurikar}, {Seifert}, {Sherbert}, {Sherwood-Taylor}, {Shih},
  {Sick}, {Silbiger}, {Singanamalla}, {Singer}, {Sladen}, {Sooley},
  {Sornarajah}, {Streicher}, {Teuben}, {Thomas}, {Tremblay}, {Turner},
  {Terr{\'o}n}, {van Kerkwijk}, {de la Vega}, {Watkins}, {Weaver}, {Whitmore},
  {Woillez}, {Zabalza}, \& {Astropy Contributors}}]{astropy}
{Astropy Collaboration}, {Price-Whelan}, A.~M., {Sip{\H o}cz}, B.~M., {et~al.}
  2018, \href{http://dx.doi.org/10.3847/1538-3881/aabc4f}{\JournalTitle{\aj},
  156, 123}

\bibitem[{{Bailes} {et~al.}(2017){Bailes}, {Jameson}, {Flynn}, {Bateman},
  {Barr}, {Bhandari}, {Bunton}, {Caleb}, {Campbell-Wilson}, \&
  {Farah}}]{2017PASA...34...45B}
{Bailes}, M., {Jameson}, A., {Flynn}, C., {et~al.} 2017,
  \href{http://dx.doi.org/10.1017/pasa.2017.39}{\JournalTitle{\pasa}, 34, e045}

\bibitem[{{Bannister} {et~al.}(2017){Bannister}, {Shannon}, {Macquart},
  {Flynn}, {Edwards}, {O'Neill}, {Os{\l}owski}, {Bailes}, {Zackay}, {Clarke},
  {D'Addario}, {Dodson}, {Hall}, {Jameson}, {Jones}, {Navarro}, {Trinh},
  {Allison}, {Anderson}, {Bell}, {Chippendale}, {Collier}, {Heald}, {Heywood},
  {Hotan}, {Lee-Waddell}, {Madrid}, {Marvil}, {McConnell}, {Popping},
  {Voronkov}, {Whiting}, {Allen}, {Bock}, {Brodrick}, {Cooray}, {DeBoer},
  {Diamond}, {Ekers}, {Gough}, {Hampson}, {Harvey-Smith}, {Hay}, {Hayman},
  {Jackson}, {Johnston}, {Koribalski}, {McClure-Griffiths}, {Mirtschin}, {Ng},
  {Norris}, {Pearce}, {Phillips}, {Roxby}, {Troup}, \& {Westmeier}}]{askap_fly}
{Bannister}, K.~W., {Shannon}, R.~M., {Macquart}, J.~P., {et~al.} 2017,
  \href{http://dx.doi.org/10.3847/2041-8213/aa71ff}{\JournalTitle{\apj}, 841,
  L12}

\bibitem[{{Bannister} {et~al.}(2019{\natexlab{a}}){Bannister}, {Deller},
  {Phillips}, {Macquart}, {Prochaska}, {Tejos}, {Ryder}, {Sadler}, {Shannon},
  {Simha}, {Day}, {McQuinn}, {North-Hickey}, {Bhandari}, {Arcus}, {Bennert},
  {Burchett}, {Bouwhuis}, {Dodson}, {Ekers}, {Farah}, {Flynn}, {James}, {Kerr},
  {Lenc}, {Mahony}, {O'Meara}, {Os{\l}owski}, {Qiu}, {Treu}, {U}, {Bateman},
  {Bock}, {Bolton}, {Brown}, {Bunton}, {Chippendale}, {Cooray}, {Cornwell},
  {Gupta}, {Hayman}, {Kesteven}, {Koribalski}, {MacLeod}, {McClure-Griffiths},
  {Neuhold}, {Norris}, {Pilawa}, {Qiao}, {Reynolds}, {Roxby}, {Shimwell},
  {Voronkov}, \& {Wilson}}]{askap_localization}
{Bannister}, K.~W., {Deller}, A.~T., {Phillips}, C., {et~al.}
  2019{\natexlab{a}}, \JournalTitle{arXiv e-prints}, arXiv:1906.11476

\bibitem[{{Bannister} {et~al.}(2019{\natexlab{b}}){Bannister}, {Deller},
  {Phillips}, {Macquart}, {Prochaska}, {Tejos}, {Ryder}, {Sadler}, {Shannon},
  {Simha}, {Day}, {McQuinn}, {North-Hickey}, {Bhandari}, {Arcus}, {Bennert},
  {Burchett}, {Bouwhuis}, {Dodson}, {Ekers}, {Farah}, {Flynn}, {James}, {Kerr},
  {Lenc}, {Mahony}, {O'Meara}, {Os{\l}owski}, {Qiu}, {Treu}, {U}, {Bateman},
  {Bock}, {Bolton}, {Brown}, {Bunton}, {Chippendale}, {Cooray}, {Cornwell},
  {Gupta}, {Hayman}, {Kesteven}, {Koribalski}, {MacLeod}, {McClure-Griffiths},
  {Neuhold}, {Norris}, {Pilawa}, {Qiao}, {Reynolds}, {Roxby}, {Shimwell},
  {Voronkov}, \& {Wilson}}]{bannister2019}
---. 2019{\natexlab{b}}, \JournalTitle{arXiv e-prints}, arXiv:1906.11476

\bibitem[{{Bates} {et~al.}(2014){Bates}, {Lorimer}, {Rane}, \&
  {Swiggum}}]{psrpoppy}
{Bates}, S.~D., {Lorimer}, D.~R., {Rane}, A., \& {Swiggum}, J. 2014,
  \href{http://dx.doi.org/10.1093/mnras/stu157}{\JournalTitle{\mnras}, 439,
  2893}

\bibitem[{{Bates} {et~al.}(2015){Bates}, {Lorimer}, {Rane}, \&
  {Swiggum}}]{2015ascl.soft01006B}
---. 2015, {PsrPopPy: Pulsar Population Modelling Programs in Python},
  Astrophysics Source Code Library,
  \href{http://arxiv.org/abs/1501.006}{{\sffamily ascl:1501.006}}

\bibitem[{{Bates} {et~al.}(2013){Bates}, {Lorimer}, \& {Verbiest}}]{bates2013}
{Bates}, S.~D., {Lorimer}, D.~R., \& {Verbiest}, J.~P.~W. 2013,
  \href{http://dx.doi.org/10.1093/mnras/stt257}{\JournalTitle{\mnras}, 431,
  1352}

\bibitem[{{Batten}(2019)}]{batten2019}
{Batten}, A. 2019,
  \href{http://dx.doi.org/10.21105/joss.01399}{\JournalTitle{The Journal of
  Open Source Software}, 4, 1399}

\bibitem[{{Beloborodov}(2019)}]{beloborodov2019}
{Beloborodov}, A.~M. 2019, \JournalTitle{arXiv e-prints},
  \href{http://arxiv.org/abs/1908.07743}{{\sffamily arXiv:1908.07743
  [astro-ph.HE]}}

\bibitem[{{Bhat} {et~al.}(2004){Bhat}, {Cordes}, {Camilo}, {Nice}, \&
  {Lorimer}}]{bhat2004}
{Bhat}, N.~D.~R., {Cordes}, J.~M., {Camilo}, F., {Nice}, D.~J., \& {Lorimer},
  D.~R. 2004, \href{http://dx.doi.org/10.1086/382680}{\JournalTitle{\apj}, 605,
  759}

\bibitem[{Bhattacharya {et~al.}(1992)Bhattacharya, Wijers, Hartman, \&
  Verbunt}]{bwhv92}
Bhattacharya, D., Wijers, R. A. M.~J., Hartman, J.~W., \& Verbunt, F. 1992,
  \JournalTitle{\aap}, 254, 198

\bibitem[{{Bhattacharya} {et~al.}(2019){Bhattacharya}, {Kumar}, \&
  {Lorimer}}]{Bhattacharya2019}
{Bhattacharya}, M., {Kumar}, P., \& {Lorimer}, D. 2019, \JournalTitle{arXiv
  e-prints}, arXiv:1902.10225

\bibitem[{{Bokeh Development Team}(2018)}]{bokeh}
{Bokeh Development Team}. 2018, Bokeh: Python library for interactive
  visualization

\bibitem[{{Caleb} {et~al.}(2016{\natexlab{a}}){Caleb}, {Flynn}, {Bailes},
  {Barr}, {Hunstead}, {Keane}, {Ravi}, \& {van Straten}}]{caleb2016}
{Caleb}, M., {Flynn}, C., {Bailes}, M., {et~al.} 2016{\natexlab{a}},
  \href{http://dx.doi.org/10.1093/mnras/stw175}{\JournalTitle{\mnras}, 458,
  708}

\bibitem[{{Caleb} {et~al.}(2019){Caleb}, {Stappers}, {Rajwade}, \&
  {Flynn}}]{caleb2019}
{Caleb}, M., {Stappers}, B.~W., {Rajwade}, K., \& {Flynn}, C. 2019,
  \href{http://dx.doi.org/10.1093/mnras/stz386}{\JournalTitle{\mnras}, 484,
  5500}

\bibitem[{{Caleb} {et~al.}(2016{\natexlab{b}}){Caleb}, {Flynn}, {Bailes},
  {Barr}, {Bateman}, {Bhandari}, {Campbell-Wilson}, {Green}, {Hunstead},
  {Jameson}, {Jankowski}, {Keane}, {Ravi}, {van Straten}, \&
  {Krishnan}}]{caleb2016utmost}
{Caleb}, M., {Flynn}, C., {Bailes}, M., {et~al.} 2016{\natexlab{b}},
  \href{http://dx.doi.org/10.1093/mnras/stw109}{\JournalTitle{Monthly Notices
  of the Royal Astronomical Society}, 458, 718}

\bibitem[{{Champion} {et~al.}(2016){Champion}, {Petroff}, {Kramer}, {Keith},
  {Bailes}, {Barr}, {Bates}, {Bhat}, {Burgay}, {Burke-Spolaor}, {Flynn},
  {Jameson}, {Johnston}, {Ng}, {Levin}, {Possenti}, {Stappers}, {van Straten},
  {Thornton}, {Tiburzi}, \& {Lyne}}]{champion}
{Champion}, D.~J., {Petroff}, E., {Kramer}, M., {et~al.} 2016,
  \href{http://dx.doi.org/10.1093/mnrasl/slw069}{\JournalTitle{\mnras}, 460,
  L30}

\bibitem[{{Chawla} {et~al.}(2017){Chawla}, {Kaspi}, {Josephy}, {Rajwade},
  {Lorimer}, {Archibald}, {DeCesar}, {Hessels}, {Kaplan}, {Karako-Argaman},
  {Kondratiev}, {Levin}, {Lynch}, {McLaughlin}, {Ransom}, {Roberts}, {Stairs},
  {Stovall}, {Swiggum}, \& {van Leeuwen}}]{chawla2017}
{Chawla}, P., {Kaspi}, V.~M., {Josephy}, A., {et~al.} 2017,
  \href{http://dx.doi.org/10.3847/1538-4357/aa7d57}{\JournalTitle{\apj}, 844,
  140}

\bibitem[{{CHIME/FRB Collaboration} {et~al.}(2018){CHIME/FRB Collaboration},
  {Amiri}, {Bandura}, {Berger}, {Bhardwaj}, {Boyce}, {Boyle}, {Brar},
  {Burhanpurkar}, {Chawla}, {Chowdhury}, {Cliche}, {Cranmer}, {Cubranic},
  {Deng}, {Denman}, {Dobbs}, {Fandino}, {Fonseca}, {Gaensler}, {Giri},
  {Gilbert}, {Good}, {Guliani}, {Halpern}, {Hinshaw}, {H{\"o}fer}, {Josephy},
  {Kaspi}, {Landecker}, {Lang}, {Liao}, {Masui}, {Mena-Parra}, {Naidu},
  {Newburgh}, {Ng}, {Patel}, {Pen}, {Pinsonneault-Marotte}, {Pleunis}, {Rafiei
  Ravandi}, {Ransom}, {Renard}, {Scholz}, {Sigurdson}, {Siegel}, {Smith},
  {Stairs}, {Tendulkar}, {Vand erlinde}, \& {Wiebe}}]{chimeoverview}
{CHIME/FRB Collaboration}, {Amiri}, M., {Bandura}, K., {et~al.} 2018,
  \href{http://dx.doi.org/10.3847/1538-4357/aad188}{\JournalTitle{\apj}, 863,
  48}

\bibitem[{{Chippendale} {et~al.}(2015){Chippendale}, {Brown}, {Beresford},
  {Hampson}, {Macleod}, {Shaw}, {Brothers}, {Cantrall}, {Forsyth}, {Hay}, \&
  {Leach}}]{chippendale}
{Chippendale}, A.~P., {Brown}, A.~J., {Beresford}, R.~J., {et~al.} 2015,
  \href{http://dx.doi.org/10.1109/ICEAA.2015.7297174}{in 2015 International
  Conference on Electromagnetics in Advanced Applications (ICEAA}, 541

\bibitem[{{Connor}(2019)}]{connor2019}
{Connor}, L. 2019,
  \href{http://dx.doi.org/10.1093/mnras/stz1666}{\JournalTitle{\mnras}, 487,
  5753}

\bibitem[{{Connor} {et~al.}(2016{\natexlab{a}}){Connor}, {Lin}, {Masui},
  {Oppermann}, {Pen}, {Peterson}, {Roman}, \& {Sievers}}]{connor2016b}
{Connor}, L., {Lin}, H.-H., {Masui}, K., {et~al.} 2016{\natexlab{a}},
  \href{http://dx.doi.org/10.1093/mnras/stw907}{\JournalTitle{\mnras}, 460,
  1054}

\bibitem[{{Connor} {et~al.}(2016{\natexlab{b}}){Connor}, {Sievers}, \&
  {Pen}}]{connor2016supernova}
{Connor}, L., {Sievers}, J., \& {Pen}, U.-L. 2016{\natexlab{b}},
  \href{http://dx.doi.org/10.1093/mnrasl/slv124}{\JournalTitle{\mnras}, 458,
  L19}

\bibitem[{{Connor} {et~al.}(2019){Connor}, {Smith}, {van Leeuwen}, {Mendrik},
  \& {Hester}}]{frb-olym}
{Connor}, L., {Smith}, K.~M., {van Leeuwen}, J., {Mendrik}, A., \& {Hester}, M.
  2019, \JournalTitle{ApJ 2019, \emph{in prep.}}

\bibitem[{{Cordes} \& {Lazio}(2002)}]{ne2001}
{Cordes}, J.~M., \& {Lazio}, T.~J.~W. 2002, \JournalTitle{ArXiv e-prints},
  astro

\bibitem[{{Cordes} \& {Lazio}(2003)}]{ne2001-2}
---. 2003, \JournalTitle{arXiv e-prints}, astro

\bibitem[{{Cordes} \& {McLaughlin}(2003)}]{cordes2003}
{Cordes}, J.~M., \& {McLaughlin}, M.~A. 2003,
  \href{http://dx.doi.org/10.1086/378231}{\JournalTitle{\apj}, 596, 1142}

\bibitem[{{Cordes} \& {Wasserman}(2016)}]{cordes2016}
{Cordes}, J.~M., \& {Wasserman}, I. 2016,
  \href{http://dx.doi.org/10.1093/mnras/stv2948}{\JournalTitle{\mnras}, 457,
  232}

\bibitem[{{Cordes} {et~al.}(2006){Cordes}, {Freire}, {Lorimer}, {Camilo},
  {Champion}, {Nice}, {Ramachandran}, {Hessels}, {Vlemmings}, {van Leeuwen},
  {Ransom}, {Bhat}, {Arzoumanian}, {McLaughlin}, {Kaspi}, {Kasian}, {Deneva},
  {Reid}, {Chatterjee}, {Han}, {Backer}, {Stairs}, {Deshpande}, \&
  {Faucher-Gigu{\`e}re}}]{cordes_palfa}
{Cordes}, J.~M., {Freire}, P.~C.~C., {Lorimer}, D.~R., {et~al.} 2006,
  \href{http://dx.doi.org/10.1086/498335}{\JournalTitle{\apj}, 637, 446}

\bibitem[{{Farah} {et~al.}(2018){Farah}, {Flynn}, {Bailes}, {Jameson},
  {Bannister}, {Barr}, {Bateman}, {Bhand ari}, {Caleb}, {Campbell-Wilson},
  {Chang}, {Deller}, {Green}, {Hunstead}, {Jankowski}, {Keane}, {Macquart},
  {M{\"o}ller}, {Onken}, {Os{\l}owski}, {Parthasarathy}, {Plant}, {Ravi},
  {Shannon}, {Tucker}, {Venkatraman Krishnan}, \& {Wolf}}]{utmost}
{Farah}, W., {Flynn}, C., {Bailes}, M., {et~al.} 2018,
  \href{http://dx.doi.org/10.1093/mnras/sty1122}{\JournalTitle{\mnras}, 478,
  1209}

\bibitem[{{Farah} {et~al.}(2019){Farah}, {Flynn}, {Bailes}, {Jameson},
  {Bateman}, {Campbell-Wilson}, {Day}, {Deller}, {Green}, {Gupta}, {Hunstead},
  {Lower}, {Os{\l}owski}, {Parthasarathy}, {Price}, {Ravi}, {Shannon},
  {Sutherland }, {Temby}, {Krishnan}, {Caleb}, {Chang}, {Cruces}, {Roy},
  {Morello}, {Onken}, {Stappers}, {Webb}, \& {Wolf}}]{farah2019five}
---. 2019,
  \href{http://dx.doi.org/10.1093/mnras/stz1748}{\JournalTitle{\mnras}, 488,
  2989}

\bibitem[{{Faucher-Gigu{\`e}re} \& {Kaspi}(2006)}]{2006ApJ...643..332F}
{Faucher-Gigu{\`e}re}, C.-A., \& {Kaspi}, V.~M. 2006,
  \href{http://dx.doi.org/10.1086/501516}{\JournalTitle{\apj}, 643, 332}

\bibitem[{{Fialkov} {et~al.}(2018){Fialkov}, {Loeb}, \&
  {Lorimer}}]{fialkov2018}
{Fialkov}, A., {Loeb}, A., \& {Lorimer}, D.~R. 2018,
  \href{http://dx.doi.org/10.3847/1538-4357/aad196}{\JournalTitle{\apj}, 863,
  132}

\bibitem[{{Ghirlanda} {et~al.}(2013){Ghirlanda}, {Ghisellini}, {Salvaterra},
  {Nava}, {Burlon}, {Tagliaferri}, {Campana}, {D'Avanzo}, \& {Melandri}}]{GRBs}
{Ghirlanda}, G., {Ghisellini}, G., {Salvaterra}, R., {et~al.} 2013,
  \href{http://dx.doi.org/10.1093/mnras/sts128}{\JournalTitle{\mnras}, 428,
  1410}

\bibitem[{{Gourdji} {et~al.}(2019){Gourdji}, {Michilli}, {Spitler}, {Hessels},
  {Seymour}, {Cordes}, \& {Chatterjee}}]{gourdji2019sample}
{Gourdji}, K., {Michilli}, D., {Spitler}, L.~G., {et~al.} 2019,
  \href{http://dx.doi.org/10.3847/2041-8213/ab1f8a}{\JournalTitle{\apjl}, 877,
  L19}

\bibitem[{Gunn \& Ostriker(1970)}]{go70}
Gunn, J.~E., \& Ostriker, J.~P. 1970, \JournalTitle{\apj}, 160, 979

\bibitem[{{Hessels} {et~al.}(2019){Hessels}, {Spitler}, {Seymour}, {Cordes},
  {Michilli}, {Lynch}, {Gourdji}, {Archibald}, {Bassa}, {Bower}, {Chatterjee},
  {Connor}, {Crawford}, {Deneva}, {Gajjar}, {Kaspi}, {Keimpema}, {Law},
  {Marcote}, {McLaughlin}, {Paragi}, {Petroff}, {Ransom}, {Scholz}, {Stappers},
  \& {Tendulkar}}]{hessels2019frb121102}
{Hessels}, J.~W.~T., {Spitler}, L.~G., {Seymour}, A.~D., {et~al.} 2019,
  \href{http://dx.doi.org/10.3847/2041-8213/ab13ae}{\JournalTitle{\apjl}, 876,
  L23}

\bibitem[{{Hewish} {et~al.}(1968){Hewish}, {Bell}, {Pilkington}, {Scott}, \&
  {Collins}}]{Hewish1968}
{Hewish}, A., {Bell}, S.~J., {Pilkington}, J.~D.~H., {Scott}, P.~F., \&
  {Collins}, R.~A. 1968,
  \href{http://dx.doi.org/10.1038/217709a0}{\JournalTitle{\nat}, 217, 709}

\bibitem[{{Hogg}(1999)}]{hogg1999}
{Hogg}, D.~W. 1999, \JournalTitle{arXiv e-prints}, astro

\bibitem[{{Hunter}(2007)}]{matplotlib}
{Hunter}, J.~D. 2007,
  \href{http://dx.doi.org/10.1109/MCSE.2007.55}{\JournalTitle{Computing in
  Science and Engineering}, 9, 90}

\bibitem[{{Inoue}(2004)}]{inoue2004}
{Inoue}, S. 2004,
  \href{http://dx.doi.org/10.1111/j.1365-2966.2004.07359.x}{\JournalTitle{\mnras},
  348, 999}

\bibitem[{{Ioka}(2003)}]{ioka2003}
{Ioka}, K. 2003, \href{http://dx.doi.org/10.1086/380598}{\JournalTitle{\apj},
  598, L79}

\bibitem[{{Izzard} \& {Halabi}(2018)}]{2018arXiv180806883I}
{Izzard}, R.~G., \& {Halabi}, G.~M. 2018, \JournalTitle{arXiv e-prints},
  arXiv:1808.06883

\bibitem[{Johnson(1994)}]{johnson1994continuous}
Johnson, N. 1994, Continuous univariate distributions (New York: Wiley)

\bibitem[{{Katz}(2014)}]{katz2014}
{Katz}, J.~I. 2014,
  \href{http://dx.doi.org/10.1103/PhysRevD.89.103009}{\JournalTitle{\prd}, 89,
  103009}

\bibitem[{{Keane}(2018)}]{keane2018}
{Keane}, E.~F. 2018,
  \href{http://dx.doi.org/10.1038/s41550-018-0603-0}{\JournalTitle{Nature
  Astronomy}, 2, 865}

\bibitem[{{Keane} \& {Petroff}(2015)}]{keane2015completeness}
{Keane}, E.~F., \& {Petroff}, E. 2015,
  \href{http://dx.doi.org/10.1093/mnras/stu2650}{\JournalTitle{\mnras}, 447,
  2852}

\bibitem[{{Keith} {et~al.}(2010){Keith}, {Jameson}, {van Straten}, {Bailes},
  {Johnston}, {Kramer}, {Possenti}, {Bates}, {Bhat}, {Burgay}, {Burke-Spolaor},
  {D'Amico}, {Levin}, {McMahon}, {Milia}, \& {Stappers}}]{htru}
{Keith}, M.~J., {Jameson}, A., {van Straten}, W., {et~al.} 2010,
  \href{http://dx.doi.org/10.1111/j.1365-2966.2010.17325.x}{\JournalTitle{\mnras},
  409, 619}

\bibitem[{{Lazarus} {et~al.}(2015){Lazarus}, {Brazier}, {Hessels},
  {Karako-Argaman}, {Kaspi}, {Lynch}, {Madsen}, {Patel}, {Ransom}, {Scholz},
  {Swiggum}, {Zhu}, {Allen}, {Bogdanov}, {Camilo}, {Cardoso}, {Chatterjee},
  {Cordes}, {Crawford}, {Deneva}, {Ferdman}, {Freire}, {Jenet}, {Knispel},
  {Lee}, {van Leeuwen}, {Lorimer}, {Lyne}, {McLaughlin}, {Siemens}, {Spitler},
  {Stairs}, {Stovall}, \& {Venkataraman}}]{lazarus}
{Lazarus}, P., {Brazier}, A., {Hessels}, J.~W.~T., {et~al.} 2015,
  \href{http://dx.doi.org/10.1088/0004-637X/812/1/81}{\JournalTitle{\apj}, 812,
  81}

\bibitem[{{Lorimer}(2011)}]{2011ascl.soft07019L}
{Lorimer}, D. 2011, {PSRPOP: Pulsar Population Modelling Programs},
  Astrophysics Source Code Library,
  \href{http://arxiv.org/abs/1107.019}{{\sffamily ascl:1107.019}}

\bibitem[{{Lorimer} {et~al.}(2007){Lorimer}, {Bailes}, {McLaughlin},
  {Narkevic}, \& {Crawford}}]{lorimer2007}
{Lorimer}, D.~R., {Bailes}, M., {McLaughlin}, M.~A., {Narkevic}, D.~J., \&
  {Crawford}, F. 2007,
  \href{http://dx.doi.org/10.1126/science.1147532}{\JournalTitle{Science}, 318,
  777}

\bibitem[{{Lorimer} {et~al.}(2013){Lorimer}, {Karastergiou}, {McLaughlin}, \&
  {Johnston}}]{lorimer2013}
{Lorimer}, D.~R., {Karastergiou}, A., {McLaughlin}, M.~A., \& {Johnston}, S.
  2013, \href{http://dx.doi.org/10.1093/mnrasl/slt098}{\JournalTitle{\mnras},
  436, L5}

\bibitem[{{Lorimer} \& {Kramer}(2012)}]{handbook}
{Lorimer}, D.~R., \& {Kramer}, M. 2012, {Handbook of Pulsar Astronomy}
  (Cambridge University Press)

\bibitem[{{Lorimer} {et~al.}(2006){Lorimer}, {Faulkner}, {Lyne}, {Manchester},
  {Kramer}, {McLaughlin}, {Hobbs}, {Possenti}, {Stairs}, {Camilo}, {Burgay},
  {D'Amico}, {Corongiu}, \& {Crawford}}]{2006MNRAS.372..777L}
{Lorimer}, D.~R., {Faulkner}, A.~J., {Lyne}, A.~G., {et~al.} 2006,
  \href{http://dx.doi.org/10.1111/j.1365-2966.2006.10887.x}{\JournalTitle{\mnras},
  372, 777}

\bibitem[{{Lu} \& {Kumar}(2018)}]{lu2018}
{Lu}, W., \& {Kumar}, P. 2018,
  \href{http://dx.doi.org/10.1093/mnras/sty716}{\JournalTitle{\mnras}, 477,
  2470}

\bibitem[{{Luo} {et~al.}(2018){Luo}, {Lee}, {Lorimer}, \& {Zhang}}]{luo2018}
{Luo}, R., {Lee}, K., {Lorimer}, D.~R., \& {Zhang}, B. 2018,
  \href{http://dx.doi.org/10.1093/mnras/sty2364}{\JournalTitle{\mnras}, 481,
  2320}

\bibitem[{Lyne {et~al.}(1985)Lyne, Manchester, \& Taylor}]{lmt85}
Lyne, A.~G., Manchester, R.~N., \& Taylor, J.~H. 1985, \JournalTitle{\mnras},
  213, 613

\bibitem[{{Maan} \& {van Leeuwen}(2017)}]{apertif}
{Maan}, Y., \& {van Leeuwen}, J. 2017, \JournalTitle{arXiv e-prints},
  arXiv:1709.06104

\bibitem[{{Macquart} \& {Ekers}(2018{\natexlab{a}})}]{jp2}
{Macquart}, J.-P., \& {Ekers}, R. 2018{\natexlab{a}},
  \href{http://dx.doi.org/10.1093/mnras/sty2083}{\JournalTitle{\mnras}, 480,
  4211}

\bibitem[{{Macquart} \& {Ekers}(2018{\natexlab{b}})}]{jp1}
{Macquart}, J.-P., \& {Ekers}, R.~D. 2018{\natexlab{b}},
  \href{http://dx.doi.org/10.1093/mnras/stx2825}{\JournalTitle{\mnras}, 474,
  1900}

\bibitem[{{Macquart} \& {Koay}(2013)}]{macquart2013}
{Macquart}, J.-P., \& {Koay}, J.~Y. 2013,
  \href{http://dx.doi.org/10.1088/0004-637X/776/2/125}{\JournalTitle{\apj},
  776, 125}

\bibitem[{{Macquart} {et~al.}(2019){Macquart}, {Shannon}, {Bannister}, {James},
  {Ekers}, \& {Bunton}}]{macquart2019spectral}
{Macquart}, J.~P., {Shannon}, R.~M., {Bannister}, K.~W., {et~al.} 2019,
  \href{http://dx.doi.org/10.3847/2041-8213/ab03d6}{\JournalTitle{\apjl}, 872,
  L19}

\bibitem[{{Madau} \& {Dickinson}(2014)}]{madau2014}
{Madau}, P., \& {Dickinson}, M. 2014,
  \href{http://dx.doi.org/10.1146/annurev-astro-081811-125615}{\JournalTitle{Annual
  Review of Astronomy and Astrophysics}, 52, 415}

\bibitem[{{Masui} {et~al.}(2015){Masui}, {Lin}, {Sievers}, {Anderson}, {Chang},
  {Chen}, {Ganguly}, {Jarvis}, {Kuo}, {Li}, {Liao}, {McLaughlin}, {Pen},
  {Peterson}, {Roman}, {Timbie}, {Voytek}, \& {Yadav}}]{gbt}
{Masui}, K., {Lin}, H.-H., {Sievers}, J., {et~al.} 2015,
  \href{http://dx.doi.org/10.1038/nature15769}{\JournalTitle{\nat}, 528, 523}

\bibitem[{{Masui} \& {Sigurdson}(2015)}]{masui2015a}
{Masui}, K.~W., \& {Sigurdson}, K. 2015,
  \href{http://dx.doi.org/10.1103/PhysRevLett.115.121301}{\JournalTitle{Physical
  Review Letters}, 115, 121301}

\bibitem[{{McQuinn}(2014)}]{mcquinn2014}
{McQuinn}, M. 2014,
  \href{http://dx.doi.org/10.1088/2041-8205/780/2/L33}{\JournalTitle{\apj},
  780, L33}

\bibitem[{{Mendrik} \& {Hester}(2019)}]{mh18}
{Mendrik}, A., \& {Hester}, M. 2019, \JournalTitle{Proc. IEEE eScience 2018,
  \emph{in prep.}}

\bibitem[{{Metzger} {et~al.}(2019){Metzger}, {Margalit}, \&
  {Sironi}}]{metzger2019}
{Metzger}, B.~D., {Margalit}, B., \& {Sironi}, L. 2019,
  \href{http://dx.doi.org/10.1093/mnras/stz700}{\JournalTitle{\mnras}, 485,
  4091}

\bibitem[{{Michilli} {et~al.}(2018){Michilli}, {Seymour}, {Hessels}, {Spitler},
  {Gajjar}, {Archibald}, {Bower}, {Chatterjee}, {Cordes}, {Gourdji}, {Heald},
  {Kaspi}, {Law}, {Sobey}, {Adams}, {Bassa}, {Bogdanov}, {Brinkman},
  {Demorest}, {Fernand ez}, {Hellbourg}, {Lazio}, {Lynch}, {Maddox}, {Marcote},
  {McLaughlin}, {Paragi}, {Ransom}, {Scholz}, {Siemion}, {Tendulkar}, {van
  Rooy}, {Wharton}, \& {Whitlow}}]{michilli}
{Michilli}, D., {Seymour}, A., {Hessels}, J.~W.~T., {et~al.} 2018,
  \href{http://dx.doi.org/10.1038/nature25149}{\JournalTitle{\nat}, 553, 182}

\bibitem[{Narayan \& Ostriker(1990)}]{no90}
Narayan, R., \& Ostriker, J.~P. 1990, \JournalTitle{\apj}, 352, 222

\bibitem[{{Niino}(2018)}]{niino2018}
{Niino}, Y. 2018,
  \href{http://dx.doi.org/10.3847/1538-4357/aab9a9}{\JournalTitle{\apj}, 858,
  4}

\bibitem[{{Oliphant}(2007)}]{scipy}
{Oliphant}, T.~E. 2007,
  \href{http://dx.doi.org/10.1109/MCSE.2007.58}{\JournalTitle{Computing in
  Science and Engineering}, 9, 10}

\bibitem[{{Oosterloo} {et~al.}(2009){Oosterloo}, {Verheijen}, {van Cappellen},
  {Bakker}, {Heald}, \& {Ivashina}}]{oosterloo}
{Oosterloo}, T., {Verheijen}, M.~A.~W., {van Cappellen}, W., {et~al.} 2009, in
  Wide Field Astronomy \&amp; Technology for the Square Kilometre Array, 70

\bibitem[{{Oppermann} {et~al.}(2016){Oppermann}, {Connor}, \&
  {Pen}}]{Oppermann16}
{Oppermann}, N., {Connor}, L.~D., \& {Pen}, U.-L. 2016,
  \href{http://dx.doi.org/10.1093/mnras/stw1401}{\JournalTitle{\mnras}, 461,
  984}

\bibitem[{{Patel} {et~al.}(2018){Patel}, {Agarwal}, {Bhardwaj}, {Boyce},
  {Brazier}, {Chatterjee}, {Chawla}, {Kaspi}, {Lorimer}, {McLaughlin},
  {Parent}, {Pleunis}, {Ransom}, {Scholz}, {Wharton}, {Zhu}, {Alam}, {Caballero
  Valdez}, {Camilo}, {Cordes}, {Crawford}, {Deneva}, {Ferdman}, {Freire},
  {Hessels}, {Nguyen}, {Stairs}, {Stovall}, \& {van Leeuwen}}]{patel}
{Patel}, C., {Agarwal}, D., {Bhardwaj}, M., {et~al.} 2018,
  \href{http://dx.doi.org/10.3847/1538-4357/aaee65}{\JournalTitle{\apj}, 869,
  181}

\bibitem[{{Petroff} {et~al.}(2019){Petroff}, {Hessels}, \&
  {Lorimer}}]{petroff_review}
{Petroff}, E., {Hessels}, J.~W.~T., \& {Lorimer}, D.~R. 2019,
  \href{http://dx.doi.org/10.1007/s00159-019-0116-6}{\JournalTitle{\aapr}, 27,
  4}

\bibitem[{{Petroff} {et~al.}(2015){Petroff}, {Johnston}, {Keane}, {van
  Straten}, {Bailes}, {Barr}, {Barsdell}, {Burke-Spolaor}, {Caleb}, {Champion},
  {Flynn}, {Jameson}, {Kramer}, {Ng}, {Possenti}, \&
  {Stappers}}]{2015MNRAS.454..457P}
{Petroff}, E., {Johnston}, S., {Keane}, E.~F., {et~al.} 2015,
  \href{http://dx.doi.org/10.1093/mnras/stv1953}{\JournalTitle{\mnras}, 454,
  457}

\bibitem[{{Petroff} {et~al.}(2016){Petroff}, {Barr}, {Jameson}, {Keane},
  {Bailes}, {Kramer}, {Morello}, {Tabbara}, \& {van Straten}}]{frbcat}
{Petroff}, E., {Barr}, E.~D., {Jameson}, A., {et~al.} 2016,
  \href{http://dx.doi.org/10.1017/pasa.2016.35}{\JournalTitle{Publications of
  the Astronomical Society of Australia}, 33, e045}

\bibitem[{{Planck Collaboration} {et~al.}(2016){Planck Collaboration}, {Ade},
  {Aghanim}, {Arnaud}, {Ashdown}, {Aumont}, {Baccigalupi}, {Banday},
  {Barreiro}, {Bartlett}, {Bartolo}, {Battaner}, {Battye}, {Benabed},
  {Beno{\^\i}t}, {Benoit-L{\'e}vy}, {Bernard}, {Bersanelli}, {Bielewicz},
  {Bock}, {Bonaldi}, {Bonavera}, {Bond}, {Borrill}, {Bouchet}, {Boulanger},
  {Bucher}, {Burigana}, {Butler}, {Calabrese}, {Cardoso}, {Catalano},
  {Challinor}, {Chamballu}, {Chary}, {Chiang}, {Chluba}, {Christensen},
  {Church}, {Clements}, {Colombi}, {Colombo}, {Combet}, {Coulais}, {Crill},
  {Curto}, {Cuttaia}, {Danese}, {Davies}, {Davis}, {de Bernardis}, {de Rosa},
  {de Zotti}, {Delabrouille}, {D{\'e}sert}, {Di Valentino}, {Dickinson},
  {Diego}, {Dolag}, {Dole}, {Donzelli}, {Dor{\'e}}, {Douspis}, {Ducout},
  {Dunkley}, {Dupac}, {Efstathiou}, {Elsner}, {En{\ss}lin}, {Eriksen},
  {Farhang}, {Fergusson}, {Finelli}, {Forni}, {Frailis}, {Fraisse},
  {Franceschi}, {Frejsel}, {Galeotta}, {Galli}, {Ganga}, {Gauthier}, {Gerbino},
  {Ghosh}, {Giard}, {Giraud-H{\'e}raud}, {Giusarma}, {Gjerl{\o}w},
  {Gonz{\'a}lez-Nuevo}, {G{\'o}rski}, {Gratton}, {Gregorio}, {Gruppuso},
  {Gudmundsson}, {Hamann}, {Hansen}, {Hanson}, {Harrison}, {Helou}, {Henrot-
  Versill{\'e}}, {Hern{\'a}ndez-Monteagudo}, {Herranz}, {Hildebrandt}, {Hivon},
  {Hobson}, {Holmes}, {Hornstrup}, {Hovest}, {Huang}, {Huffenberger}, {Hurier},
  {Jaffe}, {Jaffe}, {Jones}, {Juvela}, {Keih{\"a}nen}, {Keskitalo}, {Kisner},
  {Kneissl}, {Knoche}, {Knox}, {Kunz}, {Kurki-Suonio}, {Lagache},
  {L{\"a}hteenm{\"a}ki}, {Lamarre}, {Lasenby}, {Lattanzi}, {Lawrence}, {Leahy},
  {Leonardi}, {Lesgourgues}, {Levrier}, {Lewis}, {Liguori}, {Lilje},
  {Linden-V{\o}rnle}, {L{\'o}pez-Caniego}, {Lubin}, {Mac{\'\i}as-P{\'e}rez},
  {Maggio}, {Maino}, {Mandolesi}, {Mangilli}, {Marchini}, {Maris}, {Martin},
  {Martinelli}, {Mart{\'\i}nez-Gonz{\'a}lez}, {Masi}, {Matarrese}, {McGehee},
  {Meinhold}, {Melchiorri}, {Melin}, {Mendes}, {Mennella}, {Migliaccio},
  {Millea}, {Mitra}, {Miville-Desch{\^e}nes}, {Moneti}, {Montier}, {Morgante},
  {Mortlock}, {Moss}, {Munshi}, {Murphy}, {Naselsky}, {Nati}, {Natoli},
  {Netterfield}, {N{\o}rgaard-Nielsen}, {Noviello}, {Novikov}, {Novikov},
  {Oxborrow}, {Paci}, {Pagano}, {Pajot}, {Paladini}, {Paoletti}, {Partridge},
  {Pasian}, {Patanchon}, {Pearson}, {Perdereau}, {Perotto}, {Perrotta},
  {Pettorino}, {Piacentini}, {Piat}, {Pierpaoli}, {Pietrobon}, {Plaszczynski},
  {Pointecouteau}, {Polenta}, {Popa}, {Pratt}, {Pr{\'e}zeau}, {Prunet},
  {Puget}, {Rachen}, {Reach}, {Rebolo}, {Reinecke}, {Remazeilles}, {Renault},
  {Renzi}, {Ristorcelli}, {Rocha}, {Rosset}, {Rossetti}, {Roudier},
  {Rouill{\'e} d'Orfeuil}, {Rowan-Robinson}, {Rubi{\~n}o-Mart{\'\i}n},
  {Rusholme}, {Said}, {Salvatelli}, {Salvati}, {Sandri}, {Santos},
  {Savelainen}, {Savini}, {Scott}, {Seiffert}, {Serra}, {Shellard}, {Spencer},
  {Spinelli}, {Stolyarov}, {Stompor}, {Sudiwala}, {Sunyaev}, {Sutton},
  {Suur-Uski}, {Sygnet}, {Tauber}, {Terenzi}, {Toffolatti}, {Tomasi},
  {Tristram}, {Trombetti}, {Tucci}, {Tuovinen}, {T{\"u}rler}, {Umana},
  {Valenziano}, {Valiviita}, {Van Tent}, {Vielva}, {Villa}, {Wade}, {Wandelt},
  {Wehus}, {White}, {White}, {Wilkinson}, {Yvon}, {Zacchei}, \&
  {Zonca}}]{planck}
{Planck Collaboration}, {Ade}, P.~A.~R., {Aghanim}, N., {et~al.} 2016,
  \href{http://dx.doi.org/10.1051/0004-6361/201525830}{\JournalTitle{\aap},
  594, A13}

\bibitem[{{Platts} {et~al.}(2018){Platts}, {Weltman}, {Walters}, {Tendulkar},
  {Gordin}, \& {Kandhai}}]{frbtheories}
{Platts}, E., {Weltman}, A., {Walters}, A., {et~al.} 2018, \JournalTitle{arXiv
  e-prints}, arXiv:1810.05836

\bibitem[{{Pol} {et~al.}(2019){Pol}, {Lam}, {McLaughlin}, {Lazio}, \&
  {Cordes}}]{pol2019}
{Pol}, N., {Lam}, M.~T., {McLaughlin}, M.~A., {Lazio}, T.~J.~W., \& {Cordes},
  J.~M. 2019, \JournalTitle{arXiv e-prints}, arXiv:1903.07630

\bibitem[{{Ravi}(2019)}]{ravi2019prevalence}
{Ravi}, V. 2019,
  \href{http://dx.doi.org/10.1038/s41550-019-0831-y}{\JournalTitle{Nature
  Astronomy}, 405}

\bibitem[{{Ravi} \& {Loeb}(2019)}]{ravi2019explaining}
{Ravi}, V., \& {Loeb}, A. 2019,
  \href{http://dx.doi.org/10.3847/1538-4357/ab0748}{\JournalTitle{\apj}, 874,
  72}

\bibitem[{{Ravi} {et~al.}(2016){Ravi}, {Shannon}, {Bailes}, {Bannister},
  {Bhandari}, {Bhat}, {Burke-Spolaor}, {Caleb}, {Flynn}, {Jameson}, {Johnston},
  {Keane}, {Kerr}, {Tiburzi}, {Tuntsov}, \& {Vedantham}}]{parkesbeam}
{Ravi}, V., {Shannon}, R.~M., {Bailes}, M., {et~al.} 2016,
  \href{http://dx.doi.org/10.1126/science.aaf6807}{\JournalTitle{Science}, 354,
  1249}

\bibitem[{{Ravi} {et~al.}(2019{\natexlab{a}}){Ravi}, {Catha}, {D'Addario},
  {Djorgovski}, {Hallinan}, {Hobbs}, {Kocz}, {Kulkarni}, {Shi}, {Vedantham},
  {Weinreb}, \& {Woody}}]{dsa_localization}
{Ravi}, V., {Catha}, M., {D'Addario}, L., {et~al.} 2019{\natexlab{a}},
  \JournalTitle{arXiv e-prints}, arXiv:1907.01542

\bibitem[{{Ravi} {et~al.}(2019{\natexlab{b}}){Ravi}, {Battaglia},
  {Burke-Spolaor}, {Chatterjee}, {Cordes}, {Hallinan}, {Law}, {Lazio}, {Masui},
  {McQuinn}, {Mu{\~n}oz}, {Palliyaguru}, {Prochaska}, {Seymour}, {Vedantham},
  \& {Zheng}}]{ravi2019tomography}
{Ravi}, V., {Battaglia}, N., {Burke-Spolaor}, S., {et~al.} 2019{\natexlab{b}},
  \JournalTitle{\baas}, 51, 420

\bibitem[{{Remazeilles} {et~al.}(2015){Remazeilles}, {Dickinson}, {Banday},
  {Bigot-Sazy}, \& {Ghosh}}]{haslam408}
{Remazeilles}, M., {Dickinson}, C., {Banday}, A.~J., {Bigot-Sazy}, M.~A., \&
  {Ghosh}, T. 2015,
  \href{http://dx.doi.org/10.1093/mnras/stv1274}{\JournalTitle{\mnras}, 451,
  4311}

\bibitem[{{Romero} {et~al.}(2016){Romero}, {del Valle}, \&
  {Vieyro}}]{romero2016}
{Romero}, G.~E., {del Valle}, M.~V., \& {Vieyro}, F.~L. 2016,
  \href{http://dx.doi.org/10.1103/PhysRevD.93.023001}{\JournalTitle{\prd}, 93,
  023001}

\bibitem[{{Sanidas} {et~al.}(2019){Sanidas}, {Cooper}, {Bassa}, {Hessels},
  {Kondratiev}, {Michilli}, {Stappers}, {Tan}, {van Leeuwen}, {Cerrigone},
  {Fallows}, {Iacobelli}, {Orr{\'u}}, {Pizzo}, {Shulevski}, {Toribio}, {ter
  Veen}, {Zucca}, {Bondonneau}, {Grie{\ss}meier}, {Karastergiou}, {Kramer}, \&
  {Sobey}}]{2019A&A...626A.104S}
{Sanidas}, S., {Cooper}, S., {Bassa}, C.~G., {et~al.} 2019,
  \href{http://dx.doi.org/10.1051/0004-6361/201935609}{\JournalTitle{\aap},
  626, A104}

\bibitem[{{Scholz} {et~al.}(2016){Scholz}, {Spitler}, {Hessels}, {Chatterjee},
  {Cordes}, {Kaspi}, {Wharton}, {Bassa}, {Bogdanov}, {Camilo}, {Crawford},
  {Deneva}, {van Leeuwen}, {Lynch}, {Madsen}, {McLaughlin}, {Mickaliger},
  {Parent}, {Patel}, {Ransom}, {Seymour}, {Stairs}, {Stappers}, \&
  {Tendulkar}}]{scholz2016}
{Scholz}, P., {Spitler}, L.~G., {Hessels}, J.~W.~T., {et~al.} 2016,
  \href{http://dx.doi.org/10.3847/1538-4357/833/2/177}{\JournalTitle{\apj},
  833, 177}

\bibitem[{{Shannon} {et~al.}(2018){Shannon}, {Macquart}, {Bannister}, {Ekers},
  {James}, {Os{\l}owski}, {Qiu}, {Sammons}, {Hotan}, {Voronkov}, {Beresford},
  {Brothers}, {Brown}, {Bunton}, {Chippendale}, {Haskins}, {Leach},
  {Marquarding}, {McConnell}, {Pilawa}, {Sadler}, {Troup}, {Tuthill},
  {Whiting}, {Allison}, {Anderson}, {Bell}, {Collier}, {G{\"u}rkan}, {Heald},
  \& {Riseley}}]{askap_nature}
{Shannon}, R.~M., {Macquart}, J.~P., {Bannister}, K.~W., {et~al.} 2018,
  \href{http://dx.doi.org/10.1038/s41586-018-0588-y}{\JournalTitle{\nat}, 562,
  386}

\bibitem[{{Smits} {et~al.}(2009){Smits}, {Kramer}, {Stappers}, {Lorimer},
  {Cordes}, \& {Faulkner}}]{2009A&A...493.1161S}
{Smits}, R., {Kramer}, M., {Stappers}, B., {et~al.} 2009,
  \href{http://dx.doi.org/10.1051/0004-6361:200810383}{\JournalTitle{\aap},
  493, 1161}

\bibitem[{{Spitler} {et~al.}(2014){Spitler}, {Cordes}, {Hessels}, {Lorimer},
  {McLaughlin}, {Chatterjee}, {Crawford}, {Deneva}, {Kaspi}, {Wharton},
  {Allen}, {Bogdanov}, {Brazier}, {Camilo}, {Freire}, {Jenet},
  {Karako-Argaman}, {Knispel}, {Lazarus}, {Lee}, {van Leeuwen}, {Lynch},
  {Ransom}, {Scholz}, {Siemens}, {Stairs}, {Stovall}, {Swiggum},
  {Venkataraman}, {Zhu}, {Aulbert}, \& {Fehrmann}}]{spitler2014}
{Spitler}, L.~G., {Cordes}, J.~M., {Hessels}, J.~W.~T., {et~al.} 2014,
  \href{http://dx.doi.org/10.1088/0004-637X/790/2/101}{\JournalTitle{\apj},
  790, 101}

\bibitem[{{Spitler} {et~al.}(2016){Spitler}, {Scholz}, {Hessels}, {Bogdanov},
  {Brazier}, {Camilo}, {Chatterjee}, {Cordes}, {Crawford}, {Deneva}, {Ferdman},
  {Freire}, {Kaspi}, {Lazarus}, {Lynch}, {Madsen}, {McLaughlin}, {Patel},
  {Ransom}, {Seymour}, {Stairs}, {Stappers}, {van Leeuwen}, \& {Zhu}}]{r1}
{Spitler}, L.~G., {Scholz}, P., {Hessels}, J.~W.~T., {et~al.} 2016,
  \href{http://dx.doi.org/10.1038/nature17168}{\JournalTitle{\nat}, 531, 202}

\bibitem[{{Szary} {et~al.}(2014){Szary}, {Zhang}, {Melikidze}, {Gil}, \&
  {Xu}}]{2014ApJ...784...59S}
{Szary}, A., {Zhang}, B., {Melikidze}, G.~I., {Gil}, J., \& {Xu}, R.-X. 2014,
  \href{http://dx.doi.org/10.1088/0004-637X/784/1/59}{\JournalTitle{\apj}, 784,
  59}

\bibitem[{Taylor \& Manchester(1977)}]{tm77}
Taylor, J.~H., \& Manchester, R.~N. 1977, \JournalTitle{\apj}, 215, 885

\bibitem[{{Tendulkar} {et~al.}(2017){Tendulkar}, {Bassa}, {Cordes}, {Bower},
  {Law}, {Chatterjee}, {Adams}, {Bogdanov}, {Burke-Spolaor}, {Butler},
  {Demorest}, {Hessels}, {Kaspi}, {Lazio}, {Maddox}, {Marcote}, {McLaughlin},
  {Paragi}, {Ransom}, {Scholz}, {Seymour}, {Spitler}, {van Langevelde}, \&
  {Wharton}}]{tendulkar}
{Tendulkar}, S.~P., {Bassa}, C.~G., {Cordes}, J.~M., {et~al.} 2017,
  \href{http://dx.doi.org/10.3847/2041-8213/834/2/L7}{\JournalTitle{\apj}, 834,
  L7}

\bibitem[{{The CHIME/FRB Collaboration} {et~al.}(2019){The CHIME/FRB
  Collaboration}, {:}, {Amiri}, {Bandura}, {Bhardwaj}, {Boubel}, {Boyce},
  {Boyle}, {Brar}, {Burhanpurkar}, {Cassanelli}, {Chawla}, {Cliche},
  {Cubranic}, {Deng}, {Denman}, {Dobbs}, {Fandino}, {Fonseca}, {Gaensler},
  {Gilbert}, {Gill}, {Giri}, {Good}, {Halpern}, {Hanna}, {Hill}, {Hinshaw},
  {H{\"o}fer}, {Josephy}, {Kaspi}, {Landecker}, {Lang}, {Lin}, {Masui},
  {Mckinven}, {Mena-Parra}, {Merryfield}, {Michilli}, {Milutinovic}, {Moatti},
  {Naidu}, {Newburgh}, {Ng}, {Patel}, {Pen}, {Pinsonneault-Marotte}, {Pleunis},
  {Rafiei-Ravandi}, {Rahman}, {Ransom}, {Renard}, {Scholz}, {Shaw}, {Siegel},
  {Smith}, {Stairs}, {Tendulkar}, {Tretyakov}, {Vanderlinde}, \& {Yadav}}]{r2}
{The CHIME/FRB Collaboration}, {:}, {Amiri}, M., {et~al.} 2019,
  \JournalTitle{nature}

\bibitem[{{Thompson} {et~al.}(2017){Thompson}, {Moran}, \&
  {Swenson}}]{thompson}
{Thompson}, A.~R., {Moran}, J.~M., \& {Swenson}, George~W., J. 2017,
  {Interferometry and Synthesis in Radio Astronomy, 3rd Edition} (Krieger
  Publishing Company)

\bibitem[{{Thornton} {et~al.}(2013){Thornton}, {Stappers}, {Bailes},
  {Barsdell}, {Bates}, {Bhat}, {Burgay}, {Burke-Spolaor}, {Champion}, {Coster},
  {D'Amico}, {Jameson}, {Johnston}, {Keith}, {Kramer}, {Levin}, {Milia}, {Ng},
  {Possenti}, \& {van Straten}}]{thornton}
{Thornton}, D., {Stappers}, B., {Bailes}, M., {et~al.} 2013,
  \href{http://dx.doi.org/10.1126/science.1236789}{\JournalTitle{Science}, 341,
  53}

\bibitem[{{van der Walt} {et~al.}(2011){van der Walt}, {Colbert}, \&
  {Varoquaux}}]{numpy}
{van der Walt}, S., {Colbert}, S.~C., \& {Varoquaux}, G. 2011,
  \href{http://dx.doi.org/10.1109/MCSE.2011.37}{\JournalTitle{Computing in
  Science and Engineering}, 13, 22}

\bibitem[{{van Leeuwen} \& {Stappers}(2010)}]{ls10}
{van Leeuwen}, J., \& {Stappers}, B.~W. 2010,
  \href{http://dx.doi.org/10.1051/0004-6361/200913121}{\JournalTitle{\aap},
  509, 7}

\bibitem[{{van Leeuwen} \& {Verbunt}(2004)}]{lv04}
{van Leeuwen}, J., \& {Verbunt}, F. 2004, in IAU Symposium, Vol. 218, Young
  Neutron Stars and Their Environments, ed. {F.~Camilo \& B.~M.~Gaensler}, 41

\bibitem[{{Vedantham} {et~al.}(2016){Vedantham}, {Ravi}, {Hallinan}, \&
  {Shannon}}]{vedantham2016}
{Vedantham}, H.~K., {Ravi}, V., {Hallinan}, G., \& {Shannon}, R.~M. 2016,
  \href{http://dx.doi.org/10.3847/0004-637X/830/2/75}{\JournalTitle{\apj}, 830,
  75}

\bibitem[{{Vieyro} {et~al.}(2017){Vieyro}, {Romero}, {Bosch-Ramon}, {Marcote},
  \& {del Valle}}]{agn}
{Vieyro}, F.~L., {Romero}, G.~E., {Bosch-Ramon}, V., {Marcote}, B., \& {del
  Valle}, M.~V. 2017,
  \href{http://dx.doi.org/10.1051/0004-6361/201730556}{\JournalTitle{\aap},
  602, A64}

\bibitem[{{Walker} {et~al.}(2018){Walker}, {Ma}, \& {Breton}}]{walker2018}
{Walker}, C.~R.~H., {Ma}, Y.~Z., \& {Breton}, R.~P. 2018, \JournalTitle{arXiv
  e-prints}, arXiv:1804.01548

\bibitem[{{Wright}(2006)}]{wright2006}
{Wright}, E.~L. 2006,
  \href{http://dx.doi.org/10.1086/510102}{\JournalTitle{Publications of the
  Astronomical Society of the Pacific}, 118, 1711}

\bibitem[{{Xu} \& {Zhang}(2016)}]{xu2016}
{Xu}, S., \& {Zhang}, B. 2016,
  \href{http://dx.doi.org/10.3847/0004-637X/832/2/199}{\JournalTitle{\apj},
  832, 199}

\bibitem[{{Yang} {et~al.}(2017){Yang}, {Luo}, {Li}, \& {Zhang}}]{yang2017}
{Yang}, Y.-P., {Luo}, R., {Li}, Z., \& {Zhang}, B. 2017,
  \href{http://dx.doi.org/10.3847/2041-8213/aa6c2e}{\JournalTitle{\apj}, 839,
  L25}

\bibitem[{{Yao} {et~al.}(2017){Yao}, {Manchester}, \& {Wang}}]{ymw}
{Yao}, J.~M., {Manchester}, R.~N., \& {Wang}, N. 2017,
  \href{http://dx.doi.org/10.3847/1538-4357/835/1/29}{\JournalTitle{\apj}, 835,
  29}

\bibitem[{{Zhang}(2018)}]{zhang_dm}
{Zhang}, B. 2018,
  \href{http://dx.doi.org/10.3847/2041-8213/aae8e3}{\JournalTitle{\apj}, 867,
  L21}

\end{thebibliography}

\begin{appendix}
 \section{Beam pattern derivation}
 \label{sec:appendix:beam}
 The S/N of an FRB is partially determined by its placement in the beam of a telescope. Calculating this scaling factor, henceforth referred to as the intensity profile $I(r)$, can be a complicated task should the survey be multi-beamed. For simpler setups, or single beam surveys, the intensity profile can be approximated as a Gaussian or an Airy disk. Calculating $I(r)$ then requires just three components: the functions describing these shapes, the radial scaling of the shapes, and the maximum allowable radial offset.

 An Airy disk can be described by
 \begin{equation}
  \label{eq:airy}
  I(r) = 4 \left(\frac{J_1\left(k \sin N\right)}{k \sin N}\right)^2
 \end{equation}
 with $J_1$ the first Bessel function \citep{thompson}. The scaling factor $k$ can be expressed as follows
 \begin{equation}
  k = \frac{2 \pi a}{\lambda}
 \end{equation}
 with
 \begin{equation}
  a = \frac{A_{\rm eff}}{2}
 \end{equation}
 where $A_{\rm eff}$ is the effective area of the beam, given by
 \begin{equation}
  A_{\rm eff} = \frac{c}{\nu_{c} D \FWHM}
 \end{equation}
 with $c$ the speed of light, $\nu_{c}$ the central frequency of the survey, and $D$ a conversion factor from arcminutes to radians given by
 \begin{equation}
  D = \frac{\pi}{60 \cdot 180}
 \end{equation}
 and the `Full Width Half Maximum' $\FWHM$ given by
 \begin{equation}
  \FWHM = 2\sqrt{\frac{A_{\rm beam}}{\pi}} \cdot 60
 \end{equation}
 with the beamsize $A_{\rm beam}$ given in degrees. With
 \begin{equation}
  \lambda = \frac{c}{\nu_{c}}
 \end{equation}
 $k$ can be reduced to
 \begin{equation}
  k = \frac{\pi}{D \FWHM}
 \end{equation}

 In a similar fashion, the radial offset $N$ over an Airy disk can be given by
 \begin{equation}
  N = \frac{\FWHM}{2}\sqrt{r}M
 \end{equation}
 Obtaining a radial offset requires the diameter to be halved ($\frac{\FWHM}{2}$), and to ensure the intensity profile is sampled uniformly over a disk, a $\sqrt{r}$ is required. That leaves $M$, a scaling factor giving the maximum offset. The choice is made to set this to any of the null points of an Airy function, providing the option to choose the number of sidelobes you wish to include. To obtain the null points the following equation can be used
 \begin{equation}
  I(r) = 4 \left(\frac{J_1\left(n\right)}{n}\right)^2 = 0
 \end{equation}
 Solving for $n$, and using equation~$\ref{eq:airy}$ allows $M$ to be constructed as
 \begin{equation}
  M = \frac{2}{D \FWHM} \arcsin \left(\frac{n D \FWHM}{\pi} \right)
 \end{equation}
 Effectively the choice of the $m^{\rm th}$ $n$ allows you to choose which sidelobe you want to include. The choice is made to used the same factor for a Gaussian beam simply because the maximum offset has to be place somewhere, and equating it to $M$ allows for quick comparisons between results obtained with either the Gaussian or Airy disk.

\end{appendix}

\end{document}